\documentclass[pra,aps,superbib,citeautoscript,twocolumn]{revtex4-1}

\usepackage{amsmath,bm,mathtools}
\usepackage{amstext}
\usepackage{epsfig}
\usepackage{xcolor}
\usepackage{subfig}
\usepackage{graphicx}
\usepackage{multirow}
\usepackage{array}
\usepackage{tikz,pgfplots}

\definecolor{dgreen}{rgb}{0,.5,0}
\definecolor{dred}{rgb}{.7,.0,.0}

\newcommand{\etal}{{\it et al.}}

\def\ddroit{{\rm d}}

\begin{document}

\title{
Linear interpolation method in ensemble Kohn--Sham and range-separated  
density-functional approximations for excited states 
}

\author{
Bruno Senjean$^1$, Stefan Knecht$^2$, Hans J\o rgen Aa. Jensen$^3$, and Emmanuel Fromager$^1$
}

\affiliation{\it 
~\\
$^1$Laboratoire de Chimie Quantique,\\
Institut de Chimie, CNRS / Universit\'{e} de Strasbourg,\\
1 rue Blaise Pascal, F-67000 Strasbourg, France\\
\\
$^2$Laboratory of Physical Chemistry,\\
ETH Z\"urich,\\
Vladimir-Prelog Weg 2, CH-8093 Z{\"u}rich, Switzerland\\
\\
$^3$Department of Physics, Chemistry and Pharmacy,\\
University of Southern Denmark,\\
Campusvej 55, DK-5230 Odense M, Denmark
}


\begin{abstract}

Gross--Oliveira--Kohn density functional theory (GOK-DFT) for ensembles
is in principle very attractive, but has been hard to use in practice.
A novel, practical model based on GOK-DFT
for the calculation of electronic excitation energies is discussed.
The new model relies on two modifications of GOK-DFT:
use of range separation and use of the slope of the linearly-interpolated
ensemble energy, rather than orbital energies.
The range-separated approach is
appealing as it enables the rigorous formulation of a multi-determinant
state-averaged DFT method. In the exact theory,
the short-range density functional, that complements the long-range
wavefunction-based ensemble energy contribution, should vary with the
ensemble weights even when the density is held fixed. This
weight dependence ensures that the range-separated ensemble energy 
varies linearly with the ensemble weights. 
When the (weight-independent) ground-state short-range exchange-correlation
functional is used in this context, curvature appears thus
leading to an approximate weight-dependent excitation energy. In order
to obtain unambiguous approximate excitation energies, we propose
to interpolate linearly the ensemble energy between equiensembles. It is
shown that such a
linear
interpolation method (LIM) can be rationalized and that it effectively introduces weight dependence
effects. As proof of principle, LIM has been applied to 
He, Be, H$_2$ in both equilibrium and stretched
geometries as well as the stretched HeH$^+$ molecule. 
Very promising results have been obtained for both single (including
charge transfer) and double excitations with  
spin-independent short-range local and
semi-local functionals. Even at the Kohn--Sham
ensemble DFT level, that is recovered when the range-separation parameter
is set to zero, LIM performs better than standard
time-dependent DFT.

\end{abstract}

\maketitle


\section{Introduction}\label{intro}

The standard approach for modeling excited states in the
framework of density-functional theory (DFT) is the time-dependent (TD)
linear response regime~\cite{Casida_tddft_review_2012}. Despite its
success, due to its low
computational cost and relatively good accuracy, standard TD-DFT still suffers
from various deficiencies, one of them being the absence of multiple
excitations in the spectrum. This is directly connected with the so-called adiabatic
approximation that consists in using a 
frequency-independent exchange-correlation kernel in the linear response
equations. In order to overcome such limitations, the combination of
TD-DFT with density-matrix-~\cite{JCP12_Pernal_tddmft-srdft} or
wavefunction-based~\cite{td-hf-srdft_open_shell_Elisa,fromager2013,JCP13_Manu_soppa-srDFT}
methods by means of range separation has been
investigated recently.\\
In this work, we propose to explore a time-independent
range-separated DFT approach for excited states that is based on 
ensembles~\cite{PRA_GOK_RRprinc,PRA_GOK_EKSDFT}. One of the motivation is the need for cheaper (in
terms of computational cost) yet still reliable (in terms of accuracy) 
alternatives to standard second-order complete active space 
(CASPT2)~\cite{caspt2}
or N-electron valence state (NEVPT2)~\cite{nevpt,nevpt2-spinless} perturbation theories for modeling,
for example, photochemical
processes~\cite{JCP14_Filatov_conical_inter_REKS,Filatov-2015-Wiley}. Ensemble range-separated DFT was initially formulated
by Pastorczak \etal~\cite{PRA13_Pernal_srEDFT} The authors considered the particular case
of Boltzmann ensemble weights. The latter
were controlled by an effective temperature that can be used as a tunable
parameter, in addition to the range-separation one. As shown in
Ref.~\cite{MP14_Manu_GACE}, an exact adiabatic connection
formula can be derived for the complementary short-range
exchange-correlation energy of an ensemble. Exactly like in Kohn--Sham
(KS)
ensemble DFT~\cite{PRA_GOK_EKSDFT,PRA14_Burke_exact_GOK-DFT,JCP14_Burke_GOK-DFT_two-electron-systems}, that is also referred to as Gross--Oliveira--Kohn DFT
(GOK-DFT), the variation of the short-range exchange-correlation density functional with the
ensemble weights plays a crucial role in the calculation of
excitation energies~\cite{MP14_Manu_GACE}. So far, short-range density-functional
approximations have been developed only for the ground state, not for ensembles.
Consequently, an approximate
(weight-independent) ground-state
functional was used in Ref.~\cite{PRA13_Pernal_srEDFT}.\\ 
The weight dependence of the range-separated ensemble energy and the ambiguity in the definition of
an approximate excitation energy, that may become weight-dependent when
approximate functionals are used, will be analyzed analytically and
numerically in this work. By analogy with the fundamental gap
problem~\cite{JPCLett12_Baer_curvature_frontier_orb_energies_dft}, a simple and general linear interpolation
method is proposed and interpreted for the purpose of defining unambiguously approximate weight-independent excitation
energies. The method becomes exact if exact functionals and
wavefunctions are used. The paper is organized as follows: After a brief introduction
to ground-state range-separated DFT in Sec.~\ref{subsec:srDFT_GS},
GOK-DFT is presented in Sec.~\ref{subsec:gok-dft} and its exact range-separated extension
is formulated in Sec.~\ref{subsec:rs_edft}. The weight-independent density-functional
approximation is then discussed in detail for a two-state ensemble.
The linear interpolation method is introduced in Sec.~\ref{subsec:widfa}
and rationalized in Sec.~\ref{subsec:rationale_lim}. The
particular case of an approximate range-separated ensemble energy that
is quadratic in the ensemble weight is then treated in
Sec.~\ref{subsec:lim_for_quadratic_energy}.
Comparison is made with Ref.~\cite{PRA13_Pernal_srEDFT} and
time-dependent adiabatic linear response theory in
Sec.~\ref{subsec:compar_existing_methods}. A
generalization to higher excitations is then given in
Sec.~\ref{subsec:higherXE}.
After the computational details in Sec.~\ref{sec:comput_details}, results obtained for
He, Be, H$_2$ and HeH$^+$ are presented and discussed in
Sec.~\ref{sec:results}.
We conclude this work with a summary in Sec.~\ref{sec:conclusions}.   

\section{Theory}\label{theo}

\subsection{Range-separated density-functional theory for the ground
state}\label{subsec:srDFT_GS}

According to the Hohenberg--Kohn (HK) theorem~\cite{hktheo}, the exact
ground-state energy of an electronic system can be
obtained variationally as follows,
\begin{eqnarray}\label{eq:HK2_theo}
E_0&=&\underset{n}{\rm min}\Big\{
F[n]+
\int \ddroit{\bf r}\,v_{\rm ne}
({\bf r})\,n({\bf r})
\Big\},
\end{eqnarray}
where $v_{\rm ne}({\bf r})$ is the nuclear potential and the minimization is performed over electron densities $n({\bf r})$ that
integrate to a fixed number $N$ of electrons. The universal Levy--Lieb
(LL) functional~\cite{IJQC83_Lieb_LF_transf} equals 
\begin{eqnarray}\label{eq:universal_LL_fun}
F[n] = 
\underset{\Psi\rightarrow n}{\rm min}\langle \Psi\vert
\hat{T}+{\hat{W}_{\rm ee}}\vert\Psi\rangle,
\end{eqnarray}
where $\hat{T}$ and $\hat{W}_{\rm ee}\equiv\sum^N_{i<j}1/r_{ij}$ are the kinetic energy and
regular two-electron repulsion operators, respectively. Following
Savin~\cite{savinbook}, we consider the decomposition of the latter into
long- and short-range contributions, 
\begin{eqnarray}\label{range_sep_interaction}
{1}/{r_{12}}&=&w^{\rm lr,\mu}_{\rm ee}(r_{12})+w^{\rm sr,\mu}_{\rm
ee}(r_{12}),
\nonumber\\
w^{\rm lr,\mu}_{\rm ee}(r_{12})&=&{\rm erf}(\mu r_{12})/r_{12},
\end{eqnarray}
where $\rm erf$ is the error function and $\mu$ is a parameter in $[0,+\infty[$ that controls the range
separation, thus leading to the
partitioning
\begin{eqnarray}\label{eq:LL_range_separation}
F[n]=
F^{\rm lr,\mu}[n]+{E}^{\rm sr,\mu}_{\rm Hxc}[n],
\end{eqnarray}
with
\begin{eqnarray}\label{eq:LL_longrange}
F^{\rm lr,\mu}[n]=
\underset{\Psi\rightarrow n}{\rm min}\langle \Psi\vert
\hat{T}+{\hat{W}^{\rm lr,\mu}_{\rm ee}}\vert\Psi\rangle
,
\end{eqnarray}
and $\hat{W}^{\rm lr,\mu}_{\rm ee}\equiv\sum^N_{i<j}w^{\rm lr,\mu}_{\rm
ee}(r_{ij})$.
The complementary $\mu$-dependent short-range 
density-functional energy  ${E}^{\rm sr,\mu}_{\rm Hxc}[n]$ can be
decomposed into Hartree (H) and exchange-correlation (xc) terms, in
analogy with conventional KS-DFT, 
\begin{eqnarray} \label{srDFTfunHdef}
{E}^{\rm sr,\mu}_{\rm Hxc}[n]&=&{E}^{\rm sr,\mu}_{\rm H}[n]+{E}^{\rm
sr,\mu}_{\rm xc}[n],
\nonumber\\
E^{{\rm sr},\mu}_{\rm H}[n]&=& \frac{1}{2}\int\int {\rm d}{\mathbf{r}}{\rm d}{\mathbf{r'}}n(\mathbf{r})n(\mathbf{r'})w^{\rm sr,\mu}_{\rm ee}\left(\vert {\bf r}-{\bf r'} \vert\right).
\end{eqnarray}
Inserting
Eq.~(\ref{eq:LL_range_separation}) into
Eq.~(\ref{eq:HK2_theo}) leads to the exact expression 
\begin{eqnarray}\label{energyminpsi}
E_0 &=& 
{\displaystyle
\underset{\Psi}{\rm min}\left\{ 
\langle \Psi\vert
\hat{T}+\hat{W}^{\rm lr,\mu}_{\rm ee}+\hat{V}_{\rm ne}\vert\Psi\rangle
+E^{\rm sr,\mu}_{\rm Hxc}[n_{\Psi}]
\right\}
}
\nonumber\\
&=&
\langle \Psi_0^\mu\vert
\hat{T}+\hat{W}^{\rm lr,\mu}_{\rm ee}+\hat{V}_{\rm ne}\vert\Psi_0^\mu\rangle
+E^{\rm sr,\mu}_{\rm Hxc}[n_{\Psi_0^\mu}],
\end{eqnarray}
where $\hat{V}_{\rm ne}=
\int \ddroit{\bf r}\,v_{\rm ne}
({\bf r})\,\hat{n}({\bf r})$ and $\hat{n}(\mathbf{r})$ is the density
operator. The electron density
obtained from the trial wavefunction $\Psi$ is denoted $n_\Psi$.
The exact minimizing wavefunction
$\Psi_0^{\mu}$ in Eq.~(\ref{energyminpsi}) has the same density $n^0$ as
the physical fully-interacting ground-state wavefunction $\Psi_0$ and it fulfils the following
self-consistent equation: 
\begin{eqnarray}\label{timeindepsrdfteq}
\hat{H}^\mu[n_{\Psi_0^{\mu}}]\vert {\Psi}^{\mu}_0\rangle
={\mathcal{E}}^{\mu}_0\vert {\Psi}^{\mu}_0\rangle,
\end{eqnarray}
where
\begin{eqnarray}\label{eq:lrhamil}
&&\hat{H}^\mu[n]
=
\hat{T}+\hat{W}^{\rm lr,\mu}_{\rm
ee}+\hat{V}_{\rm ne}
+
\int \ddroit\mathbf{r}
\dfrac{\delta E^{{\rm sr,\mu}}_{\rm
Hxc}[n]}{\delta
n(\mathbf{r})}
\hat{n}(\mathbf{r})
.
\end{eqnarray}
It is readily seen
from Eqs.~(\ref{range_sep_interaction}) and (\ref{timeindepsrdfteq})
that the KS and Schr\"{o}dinger equations are recovered 
in the limit of $\mu=0$ and $\mu\rightarrow+\infty$, respectively. An exact
combination of wavefunction theory with KS-DFT is obtained in the range of 
$0<\mu<+\infty$.\\
In order to perform practical range-separated DFT calculations, local
and semi-local short-range density functionals have been developed in
recent years~\cite{toulda,erferfgaufunc,ccsrdft,Goll2009JCP}. In
addition, various wavefunction-theory-based methods have been adapted to
this context in order to describe the long-range interaction:
Hartree--Fock (HF)~\cite{Angyan2005PRA,JCPunivmu}, {second-order M\o ller-Plesset}
(MP2)~\cite{Angyan2005PRA,Fromager2008PRA,pra_MBPTn-srdft_janos}, 
the {random-phase approximation}
(RPA)~\cite{PRL_rpa-srdft_toulouse,rpa-srdft_scuseria},
{configuration interaction} (CI)~\cite{cisrdfta,cisrdftb},
{coupled-cluster} (CC)~\cite{ccsrdft},
the {multi-configurational self-consistent field}
(MCSCF)~\cite{JCPunivmu}, 
NEVPT2~\cite{nevpt2srdft}, one-electron {reduced density-matrix-functional
theory}~\cite{RohTouPer-PRA-10} (RDMFT) and the {density matrix
renormalization group} method~\cite{arXiv15_Erik_dmrg-srdft} (DMRG). In
this work, CI will be used. The orbitals, referred to as HF short-range
DFT (HF-srDFT) orbitals in the following, are generated by restricting the minimization
on the first line of Eq.~(\ref{energyminpsi}) to single determinantal wavefunctions.      
Note that, when $\mu=0$, the HF-srDFT orbitals reduce to the
conventional KS ones.\\ 
Finally, in connection with the description of excited states, let us
mention that the exact auxiliary excited states
$\{\Psi_i^{\mu}\}_{i>0}$ 
that fulfil the eigenvalue equation,
\begin{eqnarray}\label{eq:lr_eigenvalue_EX_states}
\hat{H}^\mu[n_{\Psi_0^{\mu}}]\vert {\Psi}^{\mu}_i\rangle
={\mathcal{E}}^{\mu}_i\vert {\Psi}^{\mu}_i\rangle,
\end{eqnarray}
can be used as starting
points for reaching the physical excitation
energies by means of extrapolation
techniques~\cite{JCP14_Andreas_extrapol_range_sep,RebTouTeaHelSav-JCP-14,PRA15_Elisa_extrapolation_XE_AC},
perturbation theory~\cite{RebTouTeaHelSav-MP-15}, time-dependent linear response
theory~\cite{fromager2013,JCP13_Manu_soppa-srDFT} or ensemble
range-separated
DFT~\cite{PRA13_Pernal_srEDFT,MP14_Manu_GACE}, as discussed further in
the following.\\ 

\subsection{Ensemble density-functional theory for excited
states}\label{subsec:gok-dft}

According to the GOK variational principle~\cite{PRA_GOK_RRprinc}, that
generalizes the seminal work of
Theophilou~\cite{JPC79_Theophilou_equi-ensembles} on equiensembles, the
following inequality 
\begin{eqnarray}\label{eq:GOK_VP_Tr}
E^{\mathbf{w}}\leq {\rm Tr}\left[\hat{\Gamma}^{\mathbf{w}}
\hat{H}
\right],
\end{eqnarray}
where $\hat{H}=\hat{T}+\hat{W}_{\rm ee}+\hat{V}_{\rm ne}$ and ${\rm Tr}$ denotes the trace, is fulfilled for any ensemble 
characterized by a set of
weights $\mathbf{w}\equiv(w_0,w_1,\ldots,w_{M-1})$ with 
$w_0\geq w_1\geq\ldots\geq w_{M-1}>0$ and a set of $M$ orthonormal trial
wavefunctions
$\{\overline{\Psi}_k\}_{0\leq k\leq M-1}$ from which a trial 
density matrix can be constructed: 
\begin{eqnarray}
\hat{\Gamma}^{\mathbf{w}}=\sum^{M-1}_{k=0}w_k\vert\overline{\Psi}_k\rangle\langle\overline{\Psi}_k\vert.
\end{eqnarray} 
The lower bound in Eq.~(\ref{eq:GOK_VP_Tr}) is the exact ensemble energy
\begin{eqnarray}
E^{\mathbf{w}}=\sum_{k=0}^{M-1}w_k\langle\Psi_k\vert\hat{H}\vert\Psi_k\rangle
=\sum_{k=0}^{M-1}w_kE_k,
\end{eqnarray} 
where $\Psi_k$ is the exact $k$th eigenfunction of $\hat{H}$ and
$E_0\leq E_1\leq \ldots\leq E_{M-1}$. In the following, the ensemble will 
always contain complete sets of degenerate states (referred to as "multiplets"
in Ref.~\cite{PRA_GOK_EKSDFT}). An
important consequence of the GOK principle is that the HK
theorem can be extended to ensembles of ground and excited
states~\cite{PRA_GOK_EKSDFT}, thus leading to the exact variational
expression for the ensemble energy,
\begin{eqnarray}\label{eq:EDFT_VP}
E^{\mathbf{w}}=\underset{n}{\rm
min}\left\{F^{\mathbf{w}}[n]+
\int\ddroit\mathbf{r}\;v_{\rm
ne}(\mathbf{r})n(\mathbf{r})
\right\}, 
\end{eqnarray}
where the universal LL ensemble functional is defined as follows, 
\begin{eqnarray}\label{eq:general_LL_ensfun}
F^{\mathbf{w}}[n]=\underset{\hat{\Gamma}^{\mathbf{w}}\rightarrow n}{\rm
min}\left\{{\rm Tr}\left[\hat{\Gamma}^{\mathbf{w}}(\hat{T}+\hat{W}_{\rm
ee})\right]\right\}.
\end{eqnarray}
The minimization in Eq.~(\ref{eq:general_LL_ensfun}) is restricted to ensemble density matrices with
the ensemble density $n$: 
\begin{eqnarray}
{\rm
Tr}\left[\hat{\Gamma}^{\mathbf{w}}\hat{n}(\mathbf{r})\right]=n_{\hat{\Gamma}^{\mathbf{w}}}(\mathbf{r})=n(\mathbf{r}).
\end{eqnarray}
Note that, in the following, we will use the convention
$\sum_{k=0}^{M-1}w_k=1$ so that the ensemble density integrates to the
number of electrons $N$. The minimizing density in
Eq.~(\ref{eq:EDFT_VP}) is the exact ensemble density of the physical
system 
$ 
n^{\mathbf{w}}(\mathbf{r})=\sum_{k=0}^{M-1}w_k\,n_{\Psi_k}(\mathbf{r})$.\\

In standard ensemble DFT~\cite{PRA_GOK_EKSDFT}, that is referred to
as GOK-DFT in the following, the KS partitioning of
the LL functional is used, 
\begin{eqnarray}
F^{\mathbf{w}}[n]=T_{\rm s}^{\mathbf{w}}[n]+E_{\rm Hxc}^{\mathbf{w}}[n],
\end{eqnarray}
where the non-interacting ensemble kinetic energy is defined as
\begin{eqnarray}
T_{\rm s}^{\mathbf{w}}[n]=\underset{\hat{\Gamma}^{\mathbf{w}}\rightarrow n}{\rm
min}\left\{{\rm Tr}\left[\hat{\Gamma}^{\mathbf{w}}\hat{T}\right]\right\},
\end{eqnarray}
and $E_{\rm Hxc}^{\mathbf{w}}[n]$ is the $\mathbf{w}$-dependent Hxc
functional for the ensemble,
thus leading to the exact ensemble energy expression, according to
Eq.~(\ref{eq:EDFT_VP}),
\begin{eqnarray}\label{eq:ener_GOKDFT_VP}
E^{\mathbf{w}}=\underset{\hat{\Gamma}^{\mathbf{w}}}{\rm
min}\left\{
{\rm Tr}\left[\hat{\Gamma}^{\mathbf{w}}(\hat{T}+\hat{V}_{\rm ne})\right]
+
E_{\rm Hxc}^{\mathbf{w}}[n_{\hat{\Gamma}^{\mathbf{w}}}]
\right\}. 
\end{eqnarray}
The minimizing GOK density matrix,
\begin{eqnarray}
 \hat{\Gamma}_{\rm
s}^{\mathbf{w}}=
\sum_{k=0}^{M-1}w_k
\vert\Phi^{\mathbf{w}}_k\rangle\langle\Phi^{\mathbf{w}}_k\vert,
\end{eqnarray}
reproduces the exact ensemble density of the physical system,
\begin{eqnarray}
n_{\hat{\Gamma}_{\rm
s}^{\mathbf{w}}}(\mathbf{r})=n^{\mathbf{w}}(\mathbf{r}),
\end{eqnarray}
and it fulfils the stationarity condition $\delta
\mathcal{L}^{\mathbf{w}}[\hat{\Gamma}_{\rm s}^{\mathbf{w}}]=0$ where
\begin{eqnarray}\label{eq:Lagrangian_GOK-dft}
\mathcal{L}^{\mathbf{w}}[\hat{\Gamma}^{\mathbf{w}}]&=&
{\rm Tr}\left[\hat{\Gamma}^{\mathbf{w}}(\hat{T}+\hat{V}_{\rm ne})\right]
+
E_{\rm Hxc}^{\mathbf{w}}[n_{\hat{\Gamma}^{\mathbf{w}}}]
\nonumber\\
&&+
\sum_{k=0}^{M-1}w_k\mathcal{E}^{\mathbf{w}}_k\Big(1-\langle\overline{\Psi}_k\vert\overline{\Psi}_k\rangle\Big).
\end{eqnarray}
The coefficients $\mathcal{E}^{\mathbf{w}}_k$ are Lagrange multipliers
associated with the normalization of the trial wavefunctions
$\overline{\Psi}_k$ from which the density matrix is built.  
Considering variations
$\overline{\Psi}_k\rightarrow\overline{\Psi}_k+\delta\overline{\Psi}_k$
for each individual states separately leads to the 
self-consistent GOK equations~\cite{PRA_GOK_EKSDFT}:
\begin{eqnarray}
&&\Bigg(\hat{T}+\hat{V}_{\rm ne}+
\int \ddroit\mathbf{r} \dfrac{\delta E_{\rm
Hxc}^{\mathbf{w}}[n_{\hat{\Gamma}_{\rm s}^{\mathbf{w}}}]}{\delta
n(\mathbf{r})}\hat{n}(\mathbf{r})
\Bigg)\vert \Phi^{\mathbf{w}}_k\rangle
\nonumber\\
&&=\mathcal{E}^{\mathbf{w}}_k\vert
\Phi^{\mathbf{w}}_k\rangle, \hspace{0.2cm} 0\leq k\leq M-1.
\end{eqnarray}
\subsection{Range-separated ensemble density-functional theory
}\label{subsec:rs_edft}

In analogy with ground-state range-separated DFT, the LL ensemble 
functional in Eq.~(\ref{eq:general_LL_ensfun}) can be range-separated as
follows~\cite{PRA13_Pernal_srEDFT,MP14_Manu_GACE}, 
\begin{eqnarray}\label{eq:LL_range_separation_ens}
F^{\mathbf{w}}[n]=
F^{\rm lr,\mu,\mathbf{w}}[n]+{E}^{\rm sr,\mu,\mathbf{w}}_{\rm Hxc}[n],
\end{eqnarray}
where 
\begin{eqnarray}\label{eq:general_LL_ensfun_lr}
F^{\rm lr,\mu,\mathbf{w}}[n]=\underset{\hat{\Gamma}^{\mathbf{w}}\rightarrow n}{\rm
min}\left\{{\rm Tr}\left[\hat{\Gamma}^{\mathbf{w}}(\hat{T}+\hat{W}^{\rm
lr,\mu}_{\rm
ee})\right]\right\}.
\end{eqnarray}
In the following, the short-range ensemble functional will be partitioned
into $\mathbf{w}$-independent Hartree and $\mathbf{w}$-dependent
exchange-correlation terms,
\begin{eqnarray} \label{srDFTfundef_ens}
{E}^{\rm sr,\mu,\mathbf{w}}_{\rm Hxc}[n]&=&{E}^{\rm sr,\mu}_{\rm H}[n]+{E}^{\rm
sr,\mu,\mathbf{w}}_{\rm xc}[n].
\end{eqnarray}
Note that the decomposition is arbitrary and can be exact or not,
depending on the short-range exchange-correlation functional used. In
practical calculations, local and semi-local exchange-correlation
functionals may not remove the so-called   
"ghost
interactions"~\cite{PRL02_Gross_spurious_int_EDFT,JCP14_Pernal_ghost_interaction_ensemble}
that are included into the short-range Hartree term. Such interactions are
fictitious and unwanted. Their detailed analysis, in the context of range-separated
ensemble DFT, is currently in
progress and will be presented in a separate work.\\ 
Combining Eq.~(\ref{eq:EDFT_VP}) with
Eq.~(\ref{eq:LL_range_separation_ens}) leads to the exact
range-separated ensemble
energy expression
\begin{eqnarray}\label{eq:ener_EsrDFT_VP}
E^{\mathbf{w}}=\underset{\hat{\Gamma}^{\mathbf{w}}}{\rm
min}\Big\{
&&{\rm Tr}\left[\hat{\Gamma}^{\mathbf{w}}(\hat{T}+\hat{W}^{\rm
lr,\mu}_{\rm ee}+\hat{V}_{\rm ne})\right]
\nonumber
\\
&&+
E_{\rm Hxc}^{\rm sr,\mu,\mathbf{w}}[n_{\hat{\Gamma}^{\mathbf{w}}}]
\Big\}. 
\end{eqnarray}
The minimizing long-range-interacting ensemble density matrix
$ \hat{\Gamma}
^{\mu,\mathbf{w}}=
\sum_{k=0}^{M-1}w_k
\vert\Psi^{\mu,\mathbf{w}}_k\rangle\langle\Psi^{\mu,\mathbf{w}}_k\vert$
reproduces the physical ensemble density,
\begin{eqnarray}\label{eq:dens_from_lrint_ensemble}
n_{\hat{\Gamma}
^{\mu,\mathbf{w}}}(\mathbf{r})=n^{\mathbf{w}}(\mathbf{r}),
\end{eqnarray}
and, by analogy with Eq.~(\ref{eq:Lagrangian_GOK-dft}), we conclude that it should fulfill the self-consistent equation
\begin{eqnarray}\label{eq:sc-srEDFT_eq_exact_general}
&&\Bigg(\hat{T}+\hat{W}^{\rm lr,\mu}_{\rm ee}+\hat{V}_{\rm ne}+
\int \ddroit\mathbf{r} \dfrac{\delta E_{\rm
Hxc}^{\rm sr,\mu,\mathbf{w}}[n_{\hat{\Gamma}^{\mu,\mathbf{w}}}]}{\delta
n(\mathbf{r})}\hat{n}(\mathbf{r})
\Bigg)\vert \Psi^{\mu,\mathbf{w}}_k\rangle
\nonumber\\
&&=\mathcal{E}^{\mu,\mathbf{w}}_k\vert
\Psi^{\mu,\mathbf{w}}_k\rangle, \hspace{0.2cm} 0\leq k\leq M-1.
\end{eqnarray}
Note that the Schr\"{o}dinger and GOK-DFT equations are recovered for 
$\mu\rightarrow+\infty$ and $\mu=0$, respectively.\\

In the rest of this
work we will mainly focus on ensembles consisting of two non-degenerate
states. In this case, the ensemble weights are simply equal to 
\begin{eqnarray}
w_1=w,\hspace{0.2cm} w_0=1-w,
\end{eqnarray}
where $0\leq w\leq 1/2$, and the exact ensemble energy is a linear
function of $w$,
\begin{align}
E^w=(1-w)E_0+w\,E_1.
\end{align}
Consequently, the first excitation energy $\omega=E_1-E_0$ can be
written either as a first-order derivative,
\begin{align}\label{eq:exact_XE_dEeen_over_dw}
\omega=\dfrac{\ddroit E^w}{\ddroit w},
\end{align}
or as the slope of the linear interpolation between $w=0$ and $w=1/2$,   
\begin{align}\label{eq:exactlinearitycond}
\omega=2(E^{w=1/2}-E_0).
\end{align}
Let us stress that Eqs.~(\ref{eq:exact_XE_dEeen_over_dw}) and
(\ref{eq:exactlinearitycond}) are equivalent in the exact theory. By
using
the decomposition (see Eqs.~(\ref{eq:ener_EsrDFT_VP}) and
(\ref{eq:dens_from_lrint_ensemble})) 
\begin{eqnarray}
&&E^w=
(1-w)\langle\Psi^{\mu,w}_0\vert \hat{T}+\hat{W}^{\rm lr,\mu}_{\rm
ee}+\hat{V}_{\rm ne}\vert \Psi^{\mu,w}_0\rangle
\nonumber\\
&&+
w\langle\Psi^{\mu,w}_1\vert \hat{T}+\hat{W}^{\rm lr,\mu}_{\rm
ee}+\hat{V}_{\rm ne}\vert \Psi^{\mu,w}_1\rangle
+E^{{\rm sr,\mu,}w}_{\rm Hxc}[n^w],
\end{eqnarray}
that can be rewritten in terms of the auxiliary long-range interacting
energies as follows, according to
Eq.~(\ref{eq:sc-srEDFT_eq_exact_general}),
\begin{eqnarray}\label{eq:RS_Eens_with_auxE_exact}
&&{E}^{w}=(1-w){\mathcal{E}}^{\mu,w}_0+w{\mathcal{E}}^{\mu,w}_1
\nonumber\\
&&
-
\int \ddroit\mathbf{r}
\dfrac{\delta E^{{\rm sr,\mu},w}_{\rm
Hxc}[{n}^{w}]}{\delta
n(\mathbf{r})}{n}^{w}(\mathbf{r})
+
E^{{\rm sr,\mu},w}_{\rm Hxc}[{n}^{w}],
\end{eqnarray}
where the physical ensemble density equals the auxiliary one (see
Eq.~(\ref{eq:dens_from_lrint_ensemble})),
\begin{align}
n^w(\mathbf{r})=(1-w)n_{\Psi^{\mu,w}_0}(\mathbf{r})+w\,n_{\Psi^{\mu,w}_1}(\mathbf{r}),
\end{align}
and by applying the Hellmann--Feynman theorem,
\begin{eqnarray}
\dfrac{\ddroit {\mathcal{E}}^{\mu,w}_i}{\ddroit w}=
\int \ddroit\mathbf{r}
\dfrac{\partial }{\partial w}\Bigg(\dfrac{\delta E^{{\rm sr,\mu},w}_{\rm
Hxc}[{n}^{w}]}{\delta
n(\mathbf{r})}\Bigg){n}_{{\Psi}^{\mu,w}_i}(\mathbf{r}),
\end{eqnarray}
we finally recover from Eq.~(\ref{eq:exact_XE_dEeen_over_dw}) the following expression for the
first excitation energy~\cite{MP14_Manu_GACE}, 
\begin{eqnarray}\label{eq:exactderivative_RS_energy}
\omega
&=&\mathcal{E}^{\mu,w}_1-\mathcal{E}^{\mu,w}_0+
\left.\dfrac{\partial E^{{\rm sr,\mu,}w}_{\rm
Hxc}[n]}{\partial w}\right|_{n=n^w}
\nonumber\\
&=&\Delta\mathcal{E}^{\mu,w}+\Delta_{\rm xc}^{\mu,w}
.
\end{eqnarray}
It is readily seen from Eq.~(\ref{eq:exactderivative_RS_energy}) that
the auxiliary excitation energy
$\Delta\mathcal{E}^{\mu,w}=\mathcal{E}^{\mu,w}_1-\mathcal{E}^{\mu,w}_0$
differs in principle from the physical one. They become equal when
$\mu\rightarrow+\infty$. For finite $\mu$ values, the difference is
simply expressed in terms of a derivative with respect to the ensemble
weight 
$\Delta_{\rm xc}^{\mu,w}=\left.\partial E^{{\rm sr,\mu,}w}_{\rm
xc}[n]/\partial w\right|_{n=n^w}$.
 Note that the Hartree term 
does not contribute to the second term on the right-hand side of
Eq.~(\ref{eq:exactderivative_RS_energy}) since it is, for a given
density $n$, $w$-independent (see Eq.~(\ref{srDFTfundef_ens})).  
Interestingly, when $w\rightarrow 0$, an exact expression for the
physical excitation energy is obtained in terms of the auxiliary one
that is associated with the ground-state density (see
Eq.~(\ref{eq:lr_eigenvalue_EX_states})),
\begin{eqnarray}\label{eq:exactderivative_RS_energy_w0}
\omega
&=&\mathcal{E}^{\mu}_1-\mathcal{E}^{\mu}_0+
\left.\dfrac{\partial E^{{\rm sr,\mu,}w}_{\rm
xc}[n^0]}{\partial w}\right|_{w=0}
.
\end{eqnarray}
Note also that, when $\mu=0$ and the first excitation is a one-particle--one-hole excitation (single excitation), the
GOK expression~\cite{PRA_GOK_EKSDFT} is recovered from
Eq.~(\ref{eq:exactderivative_RS_energy}),
\begin{eqnarray}\label{eq:exactderivative_GOK_DFT_Xenergy}
\omega
=\Delta\epsilon^{w}+\Delta_{\rm xc}^{w}
,
\end{eqnarray}
where $\Delta\epsilon^{w}=\varepsilon^w_1-\varepsilon^w_0$ is the
HOMO-LUMO gap for the non-interacting ensemble and  
$\Delta_{\rm xc}^{w}=\left.\partial E^{w}_{\rm
xc}[n]/\partial w\right|_{n=n^w}$. In the $w\rightarrow0$ limit, the
exact excitation energy can be expressed in terms of the KS HOMO
$\varepsilon_0$ and LUMO $\varepsilon_1$ energies as
follows,
\begin{eqnarray}\label{eq:exactderivative_KS_DFT_Xenergy}
\omega
=\varepsilon^{w\rightarrow0}_1-\varepsilon_0
,
\end{eqnarray}
where $\varepsilon^{w\rightarrow0}_1=\varepsilon_1+\Delta_{\rm xc}^{0}$.
As shown analytically by Levy~\cite{PRA_Levy_XE-N-N-1} and numerically
by Yang {\it et al.}~\cite{PRA14_Burke_exact_GOK-DFT}, $\Delta_{\rm
xc}^{0}$ corresponds to the jump in the exchange-correlation potential
when moving from $w=0$ (ground state) to $w>0$ (ensemble of ground and
excited states). This is known as the derivative discontinuity (DD) and should
not be confused with the ground-state DD that is related to ionization
energies and electron affinities, although there are distinct  
similarities at a formal level~\cite{PRL13_Kronik_grand_can_ens,JCP14_Kronik_grand_can_ens,GouTou-PRA-14}.   
Consequently, the quantity $\Delta_{\rm xc}^{\mu,w}$ introduced in
Eq.~(\ref{eq:exactderivative_RS_energy}) will be referred to in the following as
short-range DD.


\subsection{Weight-independent density-functional approximation and the
linear interpolation method}\label{subsec:widfa}

Even though an exact adiabatic-connection-based expression exists for the
short-range ensemble exchange-correlation functional (see
Eq.~(133) in Ref.~\cite{MP14_Manu_GACE}),
it has not been used yet for developing weight-dependent density-functional
approximations. Let us stress that this is still a challenge also in the
context of GOK-DFT~\cite{PRA14_Burke_exact_GOK-DFT}. A crude approximation simply consists in using the
ground-state functional~\cite{PRA13_Pernal_srEDFT},
\begin{align}
E^{{\rm sr,\mu,}w}_{\rm xc}[n]\rightarrow E^{{\rm sr,\mu}}_{\rm
xc}[n],
\end{align}
thus leading to the approximate ensemble energy expression
\begin{eqnarray}\label{rs-widfa-ens_energy}
&&\tilde{E}^{\mu,w}=
(1-w)\langle\tilde{\Psi}^{\mu,w}_0\vert \hat{T}+\hat{W}^{\rm lr,\mu}_{\rm
ee}+\hat{V}_{\rm ne}\vert \tilde{\Psi}^{\mu,w}_0\rangle
\nonumber\\
&&+
w\langle\tilde{\Psi}^{\mu,w}_1\vert \hat{T}+\hat{W}^{\rm lr,\mu}_{\rm
ee}+\hat{V}_{\rm ne}\vert \tilde{\Psi}^{\mu,w}_1\rangle
+E^{{\rm sr,\mu}}_{\rm Hxc}[\tilde{n}^{\mu,w}],
\end{eqnarray}
that may depend on both $\mu$ and $w$, and where the approximate auxiliary ensemble density equals
\begin{align}\label{eq:aux_density_widfa}
\tilde{n}^{\mu,w}(\mathbf{r})=(1-w)n_{\tilde{\Psi}^{\mu,w}_0}(\mathbf{r})+w\,n_{\tilde{\Psi}^{\mu,w}_1}(\mathbf{r}),
\end{align}
with
\begin{eqnarray}\label{eq:sc_widfa_eq}
\hat{H}^\mu[\tilde{n}^{\mu,w}]\vert \tilde{\Psi}^{\mu,w}_i\rangle
=\tilde{\mathcal{E}}^{\mu,w}_i\vert \tilde{\Psi}^{\mu,w}_i\rangle,
\hspace{0.3cm}
i=0,1.
\end{eqnarray}
In the following we refer to this approximation as {\it
weight-independent density-functional approximation} (WIDFA).
Note that, at the WIDFA level, the ground-state density-functional 
Hamiltonian $\hat{H}^\mu[n]$ (see Eq.~(\ref{eq:lrhamil})) is used. The   
auxiliary wavefunctions $\tilde{\Psi}^{\mu,w}_i$ associated with the
bi-ensemble ($0<w\leq1/2$) will therefore deviate from their
"ground-state" limits
$\Psi^\mu_i$ ($w=0$) because of the ensemble density
$\tilde{n}^{\mu,w}$ that is inserted
into the short-range Hxc potential. Note that
Eq.~(\ref{eq:sc_widfa_eq}) should be solved self-consistently. Let us
also stress that the ground-state
short-range Hxc density-functional potential 
$
{\delta E^{{\rm sr,\mu}}_{\rm
Hxc}[n^0]}/{\delta
n(\mathbf{r})}
$   
is recovered in the limit $w\rightarrow 0$,
as readily seen from Eq.~(\ref{eq:sc_widfa_eq}).
In other words, the short-range
DD is not modeled at the WIDFA level of approximation. Finally,
the exact ($\mu$-independent) ground-state energy will still be
recovered when $w\rightarrow 0$ if no
approximation is introduced in the short-range exchange-correlation
functional,
\begin{eqnarray}
\tilde{E}^{\mu,0}=E_0.
\end{eqnarray}
Obviously, the exact ensemble energy will in general not be recovered
for $w>0$. By rewriting the WIDFA ensemble energy as
\begin{eqnarray}\label{eq:RS_Eens_with_auxE}
&&\tilde{E}^{\mu,w}=(1-w)\tilde{\mathcal{E}}^{\mu,w}_0+w\tilde{\mathcal{E}}^{\mu,w}_1
\nonumber\\
&&
-
\int \ddroit\mathbf{r}
\dfrac{\delta E^{{\rm sr,\mu}}_{\rm
Hxc}[\tilde{n}^{\mu,w}]}{\delta
n(\mathbf{r})}\tilde{n}^{\mu,w}(\mathbf{r})
+
E^{{\rm sr,\mu}}_{\rm Hxc}[\tilde{n}^{\mu,w}],
\end{eqnarray}
and applying the Hellmann--Feynman theorem,
\begin{eqnarray}\label{eq:hell-Feyn_widfa}
\dfrac{\ddroit \tilde{\mathcal{E}}^{\mu,w}_i}{\ddroit w}=
\int \ddroit\mathbf{r}
\dfrac{\partial }{\partial w}\Bigg(\dfrac{\delta E^{{\rm sr,\mu}}_{\rm
Hxc}[\tilde{n}^{\mu,w}]}{\delta
n(\mathbf{r})}\Bigg){n}_{\tilde{\Psi}^{\mu,w}_i}(\mathbf{r}),
\end{eqnarray}
we see that, within WIDFA, the first-order derivative of the ensemble
energy reduces to the auxiliary excitation energy that is in principle
$w$-dependent,
\begin{eqnarray}\label{eq:exactderivative_tilde_RS_energy}
\dfrac{\ddroit\tilde{E}^{\mu,w} }{\ddroit w}
&=&\tilde{\mathcal{E}}^{\mu,w}_1-\tilde{\mathcal{E}}^{\mu,w}_0
=\Delta\tilde{\mathcal{E}}^{\mu,w}
.
\end{eqnarray}
Therefore, in practical calculations, the WIDFA ensemble energy may not
be strictly linear in $w$, as
illustrated for He in Fig.~\ref{Fig:curvature_in_He}. In the same spirit
as Ref.~\cite{JPCLett12_Baer_curvature_frontier_orb_energies_dft}, we
propose to restore the linearity by means of a simple linear interpolation
between the ground state ($w=0$) and the equiensemble ($w=1/2$), 
\begin{eqnarray}\label{eq:exact_E_overline}
\overline{E}^{\mu,w}=E_0+2w(\tilde{E}^{\mu,1/2}-E_0).
\end{eqnarray}
This approach, that will be rationalized in
Sec.~\ref{subsec:rationale_lim}, is referred to as {\it linear interpolation method} (LIM)
in the following. The approximate excitation energy is then unambiguously defined as  
\begin{align}\label{eq:lim_XE_expression}
\omega_{\rm LIM}^{\mu}=\dfrac{\ddroit \overline{E}^{\mu,w}}{\ddroit w}=
2(\tilde{E}^{\mu,1/2}-E_0).
\end{align}
Note that, according to Eq.~(\ref{eq:exactlinearitycond}), LIM becomes exact when the exact weight-dependent short-range 
exchange-correlation functional is used. By analogy with the grand
canonical
ensemble~\cite{JPCLett12_Baer_curvature_frontier_orb_energies_dft}, we
can connect the linear interpolated and curved WIDFA ensemble energies as
follows,    
\begin{eqnarray}\label{eq:def_delta_integrand}
\overline{E}^{\mu,w}=\tilde{E}^{\mu,w}+
\int^w_0 \ddroit\xi\;\Delta_{\rm
eff}^{\mu,\xi},
\end{eqnarray}
so that, according to Eqs.~(\ref{eq:exactderivative_tilde_RS_energy})
and (\ref{eq:lim_XE_expression}),
\begin{align}\label{eq:lim_XE_expression_with_effDD}
\omega_{\rm LIM}^{\mu}
=
\Delta\tilde{\mathcal{E}}^{\mu,w}
+
\Delta_{\rm eff}^{\mu,w}.
\end{align}
As readily seen from Eqs.~(\ref{eq:exactderivative_RS_energy}) and
(\ref{eq:lim_XE_expression_with_effDD}), $\Delta_{\rm eff}^{\mu,w}$
plays the role of an effective DD that corrects for the curvature of the
WIDFA ensemble energy, thus ensuring strict linearity in $w$. A
graphical representation of LIM is given in
Fig.~\ref{Fig:curvature_corr_illustration}.     

\subsection{Rationale for LIM and the effective DD}\label{subsec:rationale_lim}
The effective DD has been introduced in
Eq.~(\ref{eq:def_delta_integrand}) for the purpose of recovering an
approximate range-separated ensemble energy that is strictly linear in
$w$. This choice can be rationalized when using a
range-dependent generalized adiabatic
connection formalism for ensembles (GACE)~\cite{MP14_Manu_GACE}, where
the exact short-range ensemble potential is adjusted so that the
auxiliary ensemble density equals the (weight-independent) density
$n(\mathbf{r})$ for any weight $\xi$ and range-separation parameter $\nu$ values:
\begin{eqnarray}\label{nlacw}
&&\Bigg(\hat{T}+\hat{W}^{\rm lr,\nu}_{\rm
ee}+\int\ddroit\mathbf{r}\,v^{\nu,\xi}(\mathbf{r})\hat{n}(\mathbf{r})
\Bigg)\vert\Psi^{\nu,\xi}_i\rangle
\nonumber
\\
&&=\mathcal{E}_i^{\nu,\xi}\vert\Psi^{\nu,\xi}_i\rangle,
\hspace{0.6cm} i=0,1,
\end{eqnarray}
where 
\begin{eqnarray}\label{nlacw}
(1-\xi)n_{\Psi_0^{\nu,\xi}}(\mathbf{r})+\xi
n_{\Psi_1^{\nu,\xi}}(\mathbf{r})
=n(\mathbf{r}).
\end{eqnarray}
It was shown~\cite{MP14_Manu_GACE} that the exact short-range
ensemble exchange-correlation density-functional energy can be formally connected
with its ground-state limit ($w=0$) 
as follows, 
\begin{eqnarray}\label{eq:Exc_AC_wANDlambda_sr}
{E}^{{\rm sr},\mu,w}_{\rm xc}[n]&=&
{E}^{{\rm sr},\mu}_{\rm xc}[n]+
\int_0^w
\ddroit \xi\, \Delta^{{\rm sr},\mu,\xi}_{\rm xc}[n]
,
\end{eqnarray}
where the exact density-functional DD equals 
\begin{eqnarray}\label{dFmux_diffmuinfty}
\displaystyle 
\Delta^{{\rm sr},\mu,\xi}_{\rm xc}[n]
&=&
\Big(\mathcal{E}^{+\infty,\xi}_1-\mathcal{E}^{+\infty,\xi}_0\Big)
-
\Big(\mathcal{E}^{\mu,\xi}_1-\mathcal{E}^{\mu,\xi}_0\Big)
.
\end{eqnarray}
When rewritting the WIDFA ensemble energy in
Eq.~(\ref{rs-widfa-ens_energy}) as 
\begin{eqnarray}
\tilde{E}^{\mu,w}&=&
F^{{\rm lr},\mu,{w}}[\tilde{n}^{\mu,w}]+
E^{{\rm sr,\mu}}_{\rm Hxc}[\tilde{n}^{\mu,w}]
\nonumber\\
&&+\int\ddroit\mathbf{r}\;v_{\rm
ne}(\mathbf{r})\tilde{n}^{\mu,w}(\mathbf{r}),
\end{eqnarray}
it becomes clear, from Eqs.~(\ref{eq:def_delta_integrand}) and (\ref{eq:Exc_AC_wANDlambda_sr}), that LIM implicitly
defines an approximate weight-dependent short-range exchange-correlation functional:
\begin{eqnarray}
{E}^{{\rm sr},\mu,w}_{\rm xc}[\tilde{n}^{\mu,w}]\rightarrow
{E}^{{\rm sr},\mu}_{\rm xc}[\tilde{n}^{\mu,w}]+
\int^w_0 \ddroit\xi\;\Delta_{\rm
eff}^{\mu,\xi}.
\end{eqnarray}
In order to connect the exact DD with the effective one,  
let us consider Eq.~(\ref{dFmux_diffmuinfty}) in the
particular case $n=\tilde{n}^{\mu,w}$ and $\xi=w$, thus leading to
\begin{eqnarray}\label{dFmux_diffmuinfty_approx_dens}
\displaystyle 
\Delta^{{\rm sr},\mu,w}_{\rm xc}[\tilde{n}^{\mu,w}]
&=&
\Delta\tilde{\mathcal{E}}^{+\infty,w}
-
\Delta\tilde{\mathcal{E}}^{\mu,w}
,
\end{eqnarray}
where $\Delta\tilde{\mathcal{E}}^{+\infty,w}$ is the excitation energy
of the fully-interacting system with ensemble density
$\tilde{n}^{\mu,w}$. If the latter is a good approximation to the true
physical ensemble density ${n}^{w}$, which is the basic assumption in WIDFA, then
$\Delta\tilde{\mathcal{E}}^{+\infty,w}$ becomes $w$-independent and
equals the true physical excitation energy. As discussed previously, the
latter has various approximate expressions that all rely on various
exact expressions. Choosing the   
slope of the
linearly-interpolated WIDFA ensemble energy $\omega_{\rm LIM}^{\mu}$ is,
in principle, as relevant as other
choices. Still, the analytical derivations and numerical results presented in
the following suggest that LIM has many advantages from a practical  
point of view. By doing so, we finally recover the expression in
Eq.~(\ref{eq:lim_XE_expression_with_effDD}):
\begin{eqnarray}\label{dFmux_diffmuinfty_approx_dens_LIM}
\displaystyle 
\Delta^{{\rm sr},\mu,w}_{\rm xc}[\tilde{n}^{\mu,w}]
&\rightarrow&
\omega_{\rm LIM}^{\mu}
-
\Delta\tilde{\mathcal{E}}^{\mu,w}
.
\end{eqnarray}

\subsection{Effective DD and excitation energy for a quadratic
range-separated ensemble energy}\label{subsec:lim_for_quadratic_energy}

For analysis purposes we will approximate the WIDFA ensemble energy by
its Taylor expansion through second order in 
$w$ (around $w=0$) over the interval $[0,1/2]$,
\begin{align}\label{eq:taylorexpansionEtilde}
\tilde{E}^{\mu,w}\rightarrow\breve{E}^{\mu,w}={E}_0+
w
\tilde{E}^{\mu (1)}
+
\dfrac{w^2}{2}\tilde{E}^{\mu (2)},
\end{align}
where, according to Eqs.~(\ref{eq:lr_eigenvalue_EX_states}),
(\ref{eq:sc_widfa_eq}), (\ref{eq:hell-Feyn_widfa}) and 
(\ref{eq:exactderivative_tilde_RS_energy}), 
\begin{eqnarray}\label{eq:auxXE_w=0_quadra}
\tilde{E}^{\mu(1)}=
\left.\dfrac{\ddroit \tilde{E}^{\mu,w}}{\ddroit w}\right
|_{w=0}=
{\mathcal{E}}^{\mu}_1-{\mathcal{E}}^{\mu}_0,
\end{eqnarray}
and
\begin{eqnarray}\label{eq:2order_deriv_ensener_quadra}
&&\tilde{E}^{\mu(2)}=
\left.\dfrac{\ddroit^2 \tilde{E}^{\mu,w}}{\ddroit w^2}\right
|_{w=0}
\nonumber\\
&&=
\int \int\ddroit\mathbf{r}\ddroit\mathbf{r'}\dfrac{\delta^2E^{{\rm
sr,\mu}}_{\rm Hxc}[n^0]}{\delta n(\mathbf{r'})\delta n(\mathbf{r})}
\big(
n_{\Psi{^\mu_1}}(\mathbf{r})-n^0(\mathbf{r})\big)
\nonumber\\
&&\times \bigg(
n_{\Psi{^\mu_1}}(\mathbf{r'})-n^0(\mathbf{r'})+\left.\dfrac{\partial
n_{\tilde{\Psi}{^{\mu,w}_0}}(\mathbf{r'})}{\partial w}\right|_{w=0}
\bigg).
\end{eqnarray}
As shown in
Sec.~\ref{sec:results}, this approximation is accurate when $\mu\geq 1.0a_0^{-1}$. For
smaller $\mu$ values, and especially in the GOK-DFT limit ($\mu=0$),
the WIDFA ensemble energy is usually not quadratic in $w$.
Nevertheless, making such an approximation gives further insight
into the LIM approach, as shown in the following.
From the equiensemble energy expression 
\begin{eqnarray}\label{eq:ensembleener_onehalf}
\breve{E}^{\mu,1/2}= E_0+
\dfrac{
1}{2}
\tilde{E}^{\mu (1)}
+
\dfrac{
1}{8}
\tilde{E}^{\mu (2)},
\end{eqnarray}
and Eq.~(\ref{eq:lim_XE_expression}), we obtain the LIM excitation
energy within the quadratic approximation, that we shall refer to
as LIM2,
\begin{eqnarray}\label{eq:lim2_exp}
\omega_{{\rm LIM}2}^{\mu}
&=& 2(\breve{E}^{\mu,1/2}-E_0)\nonumber\\
&=&\tilde{E}^{\mu (1)}
+
\dfrac{
1}{4}
\tilde{E}^{\mu (2)}
,
\end{eqnarray}
thus leading to 
\begin{eqnarray}\label{eq:LM2_XE_all_orders_srK}
&&\omega_{{\rm LIM}2}^{\mu}=
{\mathcal{E}}^\mu_1
-{\mathcal{E}}^\mu_0
\nonumber\\
&&
+\dfrac{1}{4}
\int \int\ddroit\mathbf{r}\ddroit\mathbf{r'}\dfrac{\delta^2E^{{\rm
sr,\mu}}_{\rm Hxc}[n^0]}{\delta n(\mathbf{r'})\delta n(\mathbf{r})}
\big(
n_{\Psi{^\mu_1}}(\mathbf{r})-n^0(\mathbf{r})\big)
\nonumber\\
&&\times \bigg(
n_{\Psi{^\mu_1}}(\mathbf{r'})-n^0(\mathbf{r'})+\left.\dfrac{\partial
n_{\tilde{\Psi}{^{\mu,w}_0}}(\mathbf{r'})}{\partial w}\right|_{w=0}
\bigg).
\end{eqnarray}

As shown in Appendix~\ref{appendix:sc_pt}, an explicit expression for the linear response of the ground-state
density $n_{\tilde{\Psi}{^{\mu,w}_0}}$ to variations in the ensemble
weight $w$ can be obtained from self-consistent perturbation theory.
Thus we obtain the following
expansion through second order in
the short-range kernel:
\begin{eqnarray}\label{eq:LM2_XE_second_order_srK}
&&\omega_{{\rm LIM}2}^{\mu}=
{\mathcal{E}}^\mu_1
-{\mathcal{E}}^\mu_0
\nonumber\\
&&
+\dfrac{1}{4}
\int \int\ddroit\mathbf{r}\ddroit\mathbf{r'}\dfrac{\delta^2E^{{\rm
sr,\mu}}_{\rm Hxc}[n^0]}{\delta n(\mathbf{r'})\delta n(\mathbf{r})}
\big(
n_{\Psi{^\mu_1}}(\mathbf{r'})-n^0(\mathbf{r'})\big)
\nonumber\\
&&\times
\big(
n_{\Psi{^\mu_1}}(\mathbf{r})-n^0(\mathbf{r})
\big)
\nonumber\\
&&
+\dfrac{1}{2}
\int \int\int
\int\ddroit\mathbf{r_1}\ddroit\mathbf{r'_1}\ddroit\mathbf{r}\ddroit\mathbf{r'}
\dfrac{\delta^2E^{{\rm
sr,\mu}}_{\rm Hxc}[n^0]}{\delta n(\mathbf{r_1'})\delta n(\mathbf{r_1})}
\nonumber\\
&&
\times\dfrac{\delta^2E^{{\rm
sr,\mu}}_{\rm Hxc}[n^0]}{\delta n(\mathbf{r'})\delta n(\mathbf{r})}
\big(
n_{\Psi{^\mu_1}}(\mathbf{r})-n^0(\mathbf{r})\big)
\nonumber\\
&&
\times
\big(
n_{\Psi{^\mu_1}}(\mathbf{r_1'})-n^0(\mathbf{r_1'})\big)
\sum_{i\geq1}\dfrac{{n}^\mu_{0i}(\mathbf{r_1}){n}^\mu_{0i}(\mathbf{r'})}{\mathcal{E}_0^\mu-\mathcal{E}_i^\mu}
\nonumber\\
&&+\ldots
.
\end{eqnarray}
The latter expression is convenient for comparing LIM with time-dependent range-separated
DFT, as discussed further in the following.
Returning to the quadratic ensemble energy in
Eq.~(\ref{eq:taylorexpansionEtilde}), its first-order derivative equals  
\begin{eqnarray}
\dfrac{\ddroit \breve{E}^{\mu,w}}{\ddroit w}=\tilde{E}^{\mu
(1)}+w\tilde{E}^{\mu(2)},
\end{eqnarray}
thus leading to the following expression for the effective DD, according
to Eq.~(\ref{eq:lim2_exp}),
\begin{eqnarray}
\breve{\Delta}_{\rm eff}^{\mu,w}&=&\omega_{{\rm
LIM}2}^{\mu}-\dfrac{\ddroit \breve{E}^{\mu,w}}{\ddroit w}
\nonumber\\
&=& 
\Bigg(\dfrac{1}{4}-w\Bigg)\tilde{E}^{\mu (2)}.
\end{eqnarray}
In conclusion, the effective DD is expected to vanish at $w=1/4$ when the
WIDFA ensemble energy is strictly quadratic, as illustrated in
Fig.~\ref{Fig:curvature_corr_illustration}.
\subsection{Comparison with existing
methods}\label{subsec:compar_existing_methods}

\subsubsection{Excitation energies from individual densities}

Pastorczak \etal~\cite{PRA13_Pernal_srEDFT} recently proposed to compute
excitation energies as differences of total energies,
\begin{eqnarray}
\Delta E(w)=E_1(w)-E_0(w),
\end{eqnarray}
where the energy associated with the state $i$ ($i=0,1$) is obtained
from its (individual) density as follows:
\begin{eqnarray}
E_i(w)&=&
\langle\tilde{\Psi}^{\mu,w}_i\vert \hat{T}+\hat{W}^{\rm lr,\mu}_{\rm
ee}+\hat{V}_{\rm ne}\vert \tilde{\Psi}^{\mu,w}_i\rangle
\nonumber\\
&&
+E^{{\rm sr,\mu}}_{\rm Hxc}[n_{\tilde{\Psi}^{\mu,w}_i}].
\end{eqnarray}
From the Taylor expansion 
\begin{eqnarray}
\Delta E(w)=\Delta E(0)+
w\left.\dfrac{\ddroit \Delta E(w)}{\ddroit w}\right|_{w=0}
+\mathcal{O}(w^2),
\end{eqnarray}
where
\begin{eqnarray}
\Delta E(0)&=&
{\mathcal{E}}^\mu_1
-{\mathcal{E}}^\mu_0
+
E^{{\rm sr,\mu}}_{\rm Hxc}[n_{{\Psi}^{\mu}_1}]
-
E^{{\rm sr,\mu}}_{\rm Hxc}[n^0]
\nonumber\\
&&
+\int \ddroit\mathbf{r}
\dfrac{\delta E^{{\rm sr,\mu}}_{\rm
Hxc}[n^0]}{\delta
n(\mathbf{r})}
\big(
n^0(\mathbf{r})-n_{{\Psi}^{\mu}_1}(\mathbf{r})\big),
\end{eqnarray}
and, according to Eq.~(\ref{eq:hell-Feyn_widfa}), 
\begin{eqnarray}
\left.\dfrac{\ddroit \Delta E(w)}{\ddroit w}\right|_{w=0}
&=&
\int \ddroit\mathbf{r}
\Bigg(
\dfrac{\delta E^{{\rm sr,\mu}}_{\rm
Hxc}[n_{\Psi^\mu_1}]}{\delta
n(\mathbf{r})}
-\dfrac{\delta E^{{\rm sr,\mu}}_{\rm
Hxc}[n^0]}{\delta
n(\mathbf{r})}
\Bigg)
\nonumber\\
&&\times\left. \dfrac{\partial
{n}_{\tilde{\Psi}{^{\mu,w}_1}}(\mathbf{r})}{\partial w}\right |_{w=0},
\end{eqnarray}
it is readily seen that the excitation energy will vary linearly with
$w$ in the vicinity of $w=0$. Therefore, in practical calculations, an optimal value for $w$ must be determined
~\cite{PRA13_Pernal_srEDFT}. This scheme can be compared with LIM2
by expanding the excitation energy in the density
difference $n_{\Psi{^\mu_1}}(\mathbf{r})-n^0(\mathbf{r})$, thus leading
to   
\begin{eqnarray}
&&\Delta E(w)=
{\mathcal{E}}^\mu_1
-{\mathcal{E}}^\mu_0
\nonumber\\
&&
+\dfrac{1}{2}
\int \int\ddroit\mathbf{r}\ddroit\mathbf{r'}\dfrac{\delta^2E^{{\rm
sr,\mu}}_{\rm Hxc}[n^0]}{\delta n(\mathbf{r'})\delta n(\mathbf{r})}
\big(
n_{\Psi{^\mu_1}}(\mathbf{r'})-n^0(\mathbf{r'})\big)
\nonumber\\
&&\times
\big(
n_{\Psi{^\mu_1}}(\mathbf{r})-n^0(\mathbf{r})
\big)
\nonumber\\
&&
+w
\int \int\ddroit\mathbf{r}\ddroit\mathbf{r'}\dfrac{\delta^2E^{{\rm
sr,\mu}}_{\rm Hxc}[n^0]}{\delta n(\mathbf{r'})\delta n(\mathbf{r})}
\big(
n_{\Psi{^\mu_1}}(\mathbf{r'})-n^0(\mathbf{r'})\big)
\nonumber\\
&&\times
\left. \dfrac{\partial
{n}_{\tilde{\Psi}{^{\mu,\xi}_1}}(\mathbf{r})}{\partial \xi}\right |_{\xi=0}
+\ldots
\end{eqnarray}
or, equivalently,
\begin{eqnarray}
&&\Delta E(w)=
{\mathcal{E}}^\mu_1
-{\mathcal{E}}^\mu_0
\nonumber\\
&&
+\dfrac{1}{4}
\int \int\ddroit\mathbf{r}\ddroit\mathbf{r'}\dfrac{\delta^2E^{{\rm
sr,\mu}}_{\rm Hxc}[n^0]}{\delta n(\mathbf{r'})\delta n(\mathbf{r})}
\big(
n_{\Psi{^\mu_1}}(\mathbf{r'})-n^0(\mathbf{r'})\big)
\nonumber\\
&&\times
\Bigg(
n_{\Psi{^\mu_1}}(\mathbf{r})-n^0(\mathbf{r})
+
\left. \dfrac{\partial
\tilde{n}^{\mu,w,\xi}(\mathbf{r})
}{\partial \xi}
\right |_{\xi=0}
\Bigg)
\nonumber\\
&&+\ldots
\end{eqnarray}
where
\begin{eqnarray}\label{eq:ntilde_mu_w_xi}
\tilde{n}^{\mu,w,\xi}(\mathbf{r})=
(4w+\xi){n}_{\tilde{\Psi}{^{\mu,\xi}_1}}(\mathbf{r})-
\xi n_{\tilde{\Psi}{^{\mu,\xi}_0}}(\mathbf{r}).
\end{eqnarray}
This expression is recovered from the LIM2 excitation energy in
Eq.~(\ref{eq:LM2_XE_all_orders_srK}) by applying the following substitution:
\begin{eqnarray}
n_{\tilde{\Psi}{^{\mu,\xi}_0}}(\mathbf{r})\rightarrow
\tilde{n}^{\mu,w,\xi}(\mathbf{r}).
\end{eqnarray}
In other words, for a given ensemble weight $w$, the response of
$\tilde{n}^{\mu,w,\xi}$ is used rather than the
ground-state density response in the calculation of the
excitation energy $\Delta E(w)$. 
Note that integrating
$\tilde{n}^{\mu,w,\xi}$ over space gives $4wN$. Therefore, $\tilde{n}^{\mu,w,\xi}$ may be considered as a
density only when $w=1/4$. In this case, it is simply expressed as
\begin{eqnarray}\label{eq:ntilde_1over4_w_xi}
\tilde{n}^{\mu,1/4,\xi}(\mathbf{r})=
(1+\xi){n}_{\tilde{\Psi}{^{\mu,\xi}_1}}(\mathbf{r})-
\xi n_{\tilde{\Psi}{^{\mu,\xi}_0}}(\mathbf{r}),
\end{eqnarray}
and its response to changes in $\xi$ equals
\begin{eqnarray}
\left. \dfrac{\partial
\tilde{n}^{\mu,1/4,\xi}(\mathbf{r})
}{\partial \xi}
\right |_{\xi=0}
=n_{\Psi{^\mu_1}}(\mathbf{r})-n^0(\mathbf{r})+
\left. \dfrac{\partial
{n}_{\tilde{\Psi}{^{\mu,\xi}_1}}(\mathbf{r})}{\partial \xi}\right
|_{\xi=0}.
\end{eqnarray} 
Consequently, the LIM2 excitation energy can be recovered only 
if
${n}_{\tilde{\Psi}{^{\mu,\xi}_1}}=n_{\tilde{\Psi}{^{\mu,\xi}_0}}$ around
$\xi=0$, that means when the excitation energy reduces to the auxiliary
one. Note finally that the averaged density in
Eq.~(\ref{eq:ntilde_1over4_w_xi}) can be interpreted as an ensemble
density only if $-1\leq\xi\leq -1/2$. It is unclear 
if its derivative at $\xi=0$ has any physical meaning.

\subsubsection{Time-dependent adiabatic linear response
theory}\label{subsec:linear_resp}

An approximation $\tilde{\omega}$ to the first excitation energy can also be
determined from 
range-separated DFT within the adiabatic time-dependent 
linear response regime~\cite{fromager2013,JCP13_Manu_soppa-srDFT}. The associated linear response vector
$X$ fulfils 
\begin{eqnarray}\label{eq:Casidaeq_srdft}
\Big(E^{[2]\mu}_0+K_{\rm Hxc}^{\rm sr,\mu}-\tilde{\omega}
S^{[2]\mu}\Big)X=0,
\end{eqnarray}
where the long-range interacting Hessian and the metric equal
\begin{eqnarray}\label{lrHessianmu}\begin{array} {l}
E_0^{[2]\mu}= \begin{bmatrix}
[\hat{R}_{i},[\hat{H}_0^{\mu},\hat{R}^{\dagger}_{j}]]_0
&
[\hat{R}_{i},[\hat{H}_0^{\mu},\hat{R}_{j}]]_0
\\
\Big([\hat{R}_{i},[\hat{H}_0^{\mu},\hat{R}_{j}]]_0\Big)^*
&
\Big([\hat{R}_{i},[\hat{H}_0^{\mu},\hat{R}^{\dagger}_{j}]]_0\Big)^*
\\
\end{bmatrix},
\end{array}
\end{eqnarray}
and
\begin{eqnarray}\label{metricmu}\begin{array} {l}
S^{[2]\mu}= \begin{bmatrix}
[\hat{R}_{i}
,\hat{R}^{\dagger}_{j}]_0
&
[\hat{R}_{i}
,\hat{R}_{j}]_0
\\
-\Big([\hat{R}_{i}
,\hat{R}_{j}]_0
\Big)^*
&
-\Big([\hat{R}_{i}
,\hat{R}^{\dagger}_{j}]_0\Big)^*
\\
\end{bmatrix},
\end{array}
\end{eqnarray}
respectively. Short-hand notations $[\hat{A},\hat{B}]_0=\langle
\Psi_0^\mu\vert [\hat{A},\hat{B}]\vert\Psi_0^\mu\rangle$,
$\hat{H}_0^{\mu}=\hat{H}^{\mu}[n^0]$, and
$R^\dagger_i=\vert\Psi^\mu_i\rangle\langle \Psi_0^\mu\vert$ with $i>0$ have been used.
The short-range kernel matrix in Eq.~(\ref{eq:Casidaeq_srdft}) is
written as 
\begin{eqnarray}\label{eq:srkernelmatrix}
K_{\rm Hxc}^{\rm sr,\mu}=
\int \int\ddroit\mathbf{r}\ddroit\mathbf{r'}\dfrac{\delta^2E^{{\rm
sr,\mu}}_{\rm Hxc}[n^0]}{\delta n(\mathbf{r'})\delta n(\mathbf{r})}
\,n^{[1]\mu}(\mathbf{r'})\,n^{[1]\mu\dagger}(\mathbf{r}),
\end{eqnarray}
where the gradient density vector equals
\begin{eqnarray}\label{gradientdensvector}
n^{[1]\mu}(\mathbf{r})= \begin{bmatrix}
[\hat{R}_i,\hat{n}(\mathbf{r})]_0\\
[\hat{R}^\dagger_i,\hat{n}(\mathbf{r})]_0\\
\end{bmatrix}
.
\end{eqnarray}
Since we use in this section a complete basis of orthonormal $N$-electron
eigenfunctions 
$\{\Psi^\mu_k\}_{k=0,1,\ldots}$ associated with the  
unperturbed long-range interacting Hamiltonian $\hat{H}^{\mu}[n^0]$ and 
the energies $\{\mathcal{E}_k^\mu\}_{k=0,1,\ldots}$, orbital rotations do
not need to be considered, in constrast to the approximate
multi-determinant formulations
presented in Refs.~\cite{fromager2013,JCP13_Manu_soppa-srDFT}, such that matrices simply reduce to
\begin{eqnarray}\label{lrHessianmu_simplified}
E_0^{[2]\mu}&=& \begin{bmatrix}
\big(\mathcal{E}_i^\mu-\mathcal{E}_0^\mu\big)\delta_{ij}
&
0
\\
0
&
\big(\mathcal{E}_i^\mu-\mathcal{E}_0^\mu\big)\delta_{ij}
\end{bmatrix},
\nonumber\\
S^{[2]\mu}&=& \begin{bmatrix}
\delta_{ij}&
0\\
0&
-\delta_{ij}\\
\end{bmatrix},
\end{eqnarray}
and the gradient density vector becomes
\begin{eqnarray}\label{gradientdensvector_simplified}
n^{[1]\mu}(\mathbf{r})= 
\begin{bmatrix}
{n}^\mu_{0i}(\mathbf{r})\\
-{n}^\mu_{0i}(\mathbf{r})\\
\end{bmatrix}
.
\end{eqnarray}
The transition 
matrix elements associated with the density operator
${n}^\mu_{0i}(\mathbf{r})$ have already been introduced in
Eq.~(\ref{eq:Foperator}).\\

We propose to solve Eq.~(\ref{eq:Casidaeq_srdft}) by means of 
perturbation theory in order to make a comparison with LIM2. The perturbation will be the short-range kernel. 
Let us consider the auxiliary linear
response equation, 
\begin{eqnarray}\label{eq:Casidaeq_srdft_alpha}
\Big(E^{[2]\mu}_0+\alpha K_{\rm Hxc}^{\rm sr,\mu}-\omega(\alpha)
S^{[2]\mu}\Big)
X(\alpha)=0,
\end{eqnarray}
that reduces to Eq.~(\ref{eq:Casidaeq_srdft}) in the $\alpha=1$ limit,
and the perturbation expansions
\begin{eqnarray}\label{eq:exp_2ndorder_XE_rspvec}
X(\alpha)&=&X^{(0)}+\alpha X^{(1)}+\mathcal{O}(\alpha^2),
\nonumber\\
\omega(\alpha)&=&\omega^{(0)}+\alpha\omega^{(1)}+\alpha^2\omega^{(2)}+\mathcal{O}(\alpha^3).
\end{eqnarray}
Since we are here interested in the first excitation energy only, we
have
\begin{eqnarray}\label{eq:LRvec_XE_zerothorder}
X^{(0)}=
\begin{bmatrix}
1\\
0\\
\vdots\\
0
\end{bmatrix},
\hspace{0.2cm} \omega^{(0)}=
\mathcal{E}_1^\mu-\mathcal{E}_0^\mu
.
\end{eqnarray}
Inserting Eq.~(\ref{eq:exp_2ndorder_XE_rspvec}) into
Eq.~(\ref{eq:Casidaeq_srdft_alpha}) 
leads to the following excitation energy corrections through second order,  
\begin{eqnarray}\label{EXener_1st2nd}
\omega^{(1)}&=&X^{(0)\dagger} K_{\rm Hxc}^{\rm
sr,\mu}X^{(0)},
\nonumber\\
\omega^{(2)}&=&X^{(0)\dagger} K_{\rm Hxc}^{\rm
sr,\mu}X^{(1)},
\end{eqnarray}
where the intermediate normalization condition
$X(\alpha)^\dagger S^{[2]\mu} X^{(0)}=1$ has been used,
and
\begin{eqnarray}
\Big(E^{[2]\mu}_0-\omega^{(0)}
S^{[2]\mu}\Big)
X^{(1)}&=&-K_{\rm Hxc}^{\rm
sr,\mu}X^{(0)}
\nonumber\\
&&+\omega^{(1)}S^{[2]\mu}X^{(0)}.
\end{eqnarray}
According to Eqs.~(\ref{eq:srkernelmatrix}),
(\ref{gradientdensvector_simplified}) and (\ref{eq:LRvec_XE_zerothorder}), the first-order corrections to the excitation energy
and the linear response vector become
\begin{eqnarray}\label{eq:XE_1st_corr}
\omega^{(1)}=
\int \int\ddroit\mathbf{r}\ddroit\mathbf{r'}\dfrac{\delta^2E^{{\rm
sr,\mu}}_{\rm Hxc}[n^0]}{\delta n(\mathbf{r'})\delta n(\mathbf{r})}
{n}^\mu_{01}(\mathbf{r'})
{n}^\mu_{01}(\mathbf{r})
,
\end{eqnarray}
and
\begin{eqnarray}\label{lrvec_1storder}
&&X^{(1)}=-
\int \int\ddroit\mathbf{r}\ddroit\mathbf{r'}
\dfrac{\delta^2E^{{\rm
sr,\mu}}_{\rm Hxc}[n^0]}{\delta n(\mathbf{r'})\delta n(\mathbf{r})}
\,{n}^\mu_{01}(\mathbf{r})
\nonumber\\
&&
\times 
\Big(E^{[2]\mu}_0-\omega^{(0)}
S^{[2]\mu}\Big)^{-1}
\Big(
n^{[1]\mu}(\mathbf{r'})-
{n}^\mu_{01}(\mathbf{r'})X^{(0)}
\Big),
\end{eqnarray}
respectively.
Combining Eq.~(\ref{eq:srkernelmatrix}) with Eqs.~(\ref{EXener_1st2nd})
and (\ref{lrvec_1storder}) leads to the following expression for the
second-order correction to the excitation energy:
\begin{eqnarray}\label{EXener_2ndcorr}
&&\omega^{(2)}=\int \int\int
\int\ddroit\mathbf{r_1}\ddroit\mathbf{r'_1}\ddroit\mathbf{r}\ddroit\mathbf{r'}
\dfrac{\delta^2E^{{\rm
sr,\mu}}_{\rm Hxc}[n^0]}{\delta n(\mathbf{r_1'})\delta n(\mathbf{r_1})}
\nonumber\\
&&
\times\dfrac{\delta^2E^{{\rm
sr,\mu}}_{\rm Hxc}[n^0]}{\delta n(\mathbf{r'})\delta n(\mathbf{r})}
{n}^\mu_{01}(\mathbf{r}){n}^\mu_{01}(\mathbf{r_1'})
\Bigg(
\sum_{i>1}\dfrac{{n}^\mu_{0i}(\mathbf{r_1}){n}^\mu_{0i}(\mathbf{r'})}{\mathcal{E}_1^\mu-\mathcal{E}_i^\mu}
\nonumber\\
&&+\sum_{i\geq1}\dfrac{{n}^\mu_{0i}(\mathbf{r_1}){n}^\mu_{0i}(\mathbf{r'})}{2\mathcal{E}_0^\mu-\mathcal{E}_i^\mu-\mathcal{E}_1^\mu}
\Bigg).
\end{eqnarray}
The second summation in Eq.~(\ref{EXener_2ndcorr}) is related to de-excitations. 
Within the Tamm--Dancoff approximation the latter will be dropped, thus
leading to the following expansion through second order,   
according to Eqs.~(\ref{eq:LRvec_XE_zerothorder}) and
(\ref{eq:XE_1st_corr}), 
\begin{eqnarray}\label{eq:EXener_upto2ndorder}
&&\tilde{\omega}=
\mathcal{E}_1^\mu-\mathcal{E}_0^\mu
+
\int \int\ddroit\mathbf{r}\ddroit\mathbf{r'}\dfrac{\delta^2E^{{\rm
sr,\mu}}_{\rm Hxc}[n^0]}{\delta n(\mathbf{r'})\delta n(\mathbf{r})}
{n}^\mu_{01}(\mathbf{r'})
{n}^\mu_{01}(\mathbf{r})
\nonumber
\\
&&+
\int \int\int
\int\ddroit\mathbf{r_1}\ddroit\mathbf{r'_1}\ddroit\mathbf{r}\ddroit\mathbf{r'}
\dfrac{\delta^2E^{{\rm
sr,\mu}}_{\rm Hxc}[n^0]}{\delta n(\mathbf{r_1'})\delta n(\mathbf{r_1})}
\nonumber\\
&&
\times\dfrac{\delta^2E^{{\rm
sr,\mu}}_{\rm Hxc}[n^0]}{\delta n(\mathbf{r'})\delta n(\mathbf{r})}
{n}^\mu_{01}(\mathbf{r}){n}^\mu_{01}(\mathbf{r_1'})
\sum_{i>1}\dfrac{{n}^\mu_{0i}(\mathbf{r_1}){n}^\mu_{0i}(\mathbf{r'})}{\mathcal{E}_1^\mu-\mathcal{E}_i^\mu}
\nonumber\\
&&+\ldots
\end{eqnarray}
A direct comparison can then be made with the LIM2 excitation energy 
in Eq.~(\ref{eq:LM2_XE_second_order_srK}). Thus we conclude that LIM2 can be 
recovered through first and second orders in the short-range kernel from
adiabatic time-dependent range-separated DFT 
by applying, 
within the Tamm--Dancoff
approximation, 
the following substitutions,
\begin{eqnarray}
{n}^\mu_{01}(\mathbf{r})\rightarrow\dfrac{1}{2}
\Big(
n_{\Psi{^\mu_1}}(\mathbf{r})-n^0(\mathbf{r})\Big),
\end{eqnarray}
and
\begin{eqnarray}
\sum_{i>1}\dfrac{{n}^\mu_{0i}(\mathbf{r_1}){n}^\mu_{0i}(\mathbf{r'})}{\mathcal{E}_1^\mu-\mathcal{E}_i^\mu}
\rightarrow
2\sum_{i\geq1}\dfrac{{n}^\mu_{0i}(\mathbf{r_1}){n}^\mu_{0i}(\mathbf{r'})}{\mathcal{E}_0^\mu-\mathcal{E}_i^\mu}
,
\end{eqnarray}
respectively.

\subsection{Generalization to higher excitations}\label{subsec:higherXE}

Following Gross~{\it et al}.~\cite{PRA_GOK_EKSDFT}, we introduce the generalized $w$-dependent ensemble energy
\begin{eqnarray}
\displaystyle E^w_{I} = \dfrac{1-wg_I}{\displaystyle M_{I-1}}\times\Bigg(\sum^{I-1}_{K=0}g_KE_K\Bigg) + wg_IE_I,
\end{eqnarray}
that is associated with the following ensemble weights,
\begin{eqnarray}
w_k= 
\begin{dcases}
    \dfrac{1-wg_I}{\displaystyle M_{I-1}} & 0\leq k\leq M_{I-1}-1,\\
    w              & M_{I-1} \leq k\leq M_I-1, 
\end{dcases}
\end{eqnarray}
with \begin{eqnarray}\label{eq:w_values_general_case}
0\leq w\leq\dfrac{1}{M_I},\nonumber\\
\displaystyle M_{I}=\sum^{I}_{L=0}g_L,
\end{eqnarray}
and $E_0<E_1<\ldots<E_I$ are the $I+1$ lowest energies with degeneracies
$\{g_L\}_{0\leq L\leq I}$. 
In the exact theory, the ensemble energy is linear in $w$ with slope 
\begin{eqnarray}\label{eq:GEens_deriv}
\dfrac{{\rm d}E^{w}_{I}}{{\rm d}w}=g_IE_I-\dfrac{g_I}{\displaystyle M_{I-1}}\Bigg(\sum^{I-1}_{K=0}g_KE_K\Bigg),
\end{eqnarray}
thus leading to the following expression for the exact $I$th excitation energy  
\begin{eqnarray}
\omega_I&=&E_I-E_0
\nonumber
\\
&=& \dfrac{1}{g_I}\dfrac{{\rm d}E^{w}_{I}}{{\rm d}w}+\dfrac{1}{M_{I-1}}\displaystyle\sum^{I-1}_{K=1}g_K\omega_K.
\end{eqnarray}
The LIM excitation energy, that has been introduced in Eq.~(\ref{eq:lim_XE_expression}) for
non-degenerate ground and first-excited states, can therefore be generalized
by substituting the approximate first-order derivative (that may be both $\mu$- and
 $w$-dependent) with its linear-interpolated value over the segment $[0,1/M_I]$,   
\begin{eqnarray}
\dfrac{{\rm d}\tilde{E}^{\mu,w}_{I}}{{\rm d}w}\rightarrow M_I\Big(\tilde{E}^{\mu,1/M_I}_{I}
-\tilde{E}^{\mu,0}_{I}
\Big),
\end{eqnarray}
so that the $I$th LIM excitation energy can be defined as 
\begin{eqnarray}\label{eq:limXE_general_exp}
\omega_{{\rm LIM},I}^{\mu}&=&\dfrac{M_I}{g_I}\Big(\tilde{E}^{\mu,1/M_I}_{I}-\tilde{E}^{\mu,1/M_{I-1}}_{I-1}\Big)
\nonumber\\
&&+\dfrac{1}{M_{I-1}}\sum^{I-1}_{K=1}g_K\omega_{{\rm LIM},K}^{\mu},
\end{eqnarray}
where the equality $\tilde{E}^{\mu,1/M_{I-1}}_{I-1}=\tilde{E}^{\mu,0}_{I}$ has been used.
In other words, LIM simply consists in interpolating linearly the ensemble
energy between equiensembles that are described at the WIDFA level of
approximation.

\section{Computational details}\label{sec:comput_details}

Eqs.~(\ref{eq:sc_widfa_eq}) and (\ref{eq:lim_XE_expression}) as well as
their generalizations to any ensemble of ground- and excited states (see
Eq.~(\ref{eq:limXE_general_exp})) have been implemented in a development
version of the {DALTON} program package~\cite{daltonpaper,DALTON}.
For simplicity, we considered spin-projected (singlet) ensembles only. In
the latter case, the GOK variational principle is simply formulated in
the space of singlet states~\cite{PRA14_Burke_exact_GOK-DFT}. In practice, both singlet and triplet
states have been computed but, for the latter (that can be identified
easily in
a CI calculation), the ensemble weight has
been set to zero.   
Both spin-independent short-range local density~\cite{savinbook,toulda} (srLDA) and
Perdew-Burke-Ernzerhof-type~\cite{ccsrdft} (srPBE) approximations have been used. Basis sets are
aug-cc-pVQZ~\cite{augQZbe,pVTZ_he}. Orbitals relaxation and long-range
correlation effects have been treated self-consistently at the full
CI level (FCI) in the basis of the (ground-state)
HF-srDFT orbitals. For Be, the $1s$ orbitals were
kept inactive. Indeed, in the standard wavefunction limit ($\mu\rightarrow+\infty$),
deviations from time-dependent CC with singles and doubles
(TD-CCSD) excitation energies are 0.4 and 2.0 m$E_h$ for the $2s\rightarrow
3s$ and $(2s)^2\rightarrow(2p)^2$ excitations, respectively. Comparisons
are made with standard TD-DFT using LDA~\cite{dft-Vosko-CJP1980a},
PBE~\cite{dft-Perdew-PRL1996a} and the Coulomb attenuated
Becke three-parameter Lee-Yang-Parr~\cite{dft-Yanai-CPL2004a}(CAM-B3LYP)
functionals. We investigated the following ensembles consisting of two
singlet states: $\{1^1S,2^1S\}$ for He and Be,
$\{1^1\Sigma^+,2^1\Sigma^+\}$ for the stretched HeH$^+$ molecule and
$\{1^1\Sigma_g^+,2^1\Sigma_g^+\}$ for H$_2$ at equilibrium and stretched
geometries. For Be, the four-state ensemble $\{1^1S,2^1S,1^1D\}$ in
$A_g$ symmetry ($1^1D$ is doubly degenerate) has also been considered in order to
compute the $1^1S\rightarrow1^1D$ excitation energy.    

\section{Results and discussion}\label{sec:results}

\subsection{Effective derivative discontinuities}

\subsubsection{GOK-DFT results ($\mu=0$) for He}\label{subsubsec:He_GOK}

Let us first focus on the GOK-LDA results ($\mu=0$ limit) obtained for
He. As shown in the top left-hand panel of Fig.~\ref{Fig:He_w_curves},
the variation of the auxiliary excitation energy with $w$ is very
similar to the one obtained at the quasi-LDA (qLDA) level by Yang {\it et al.}
(see Fig.~11 in Ref.~\cite{PRA14_Burke_exact_GOK-DFT}). An interesting
feature, observed with both methods, is the minimum around $w=0.01$. 
The derivation of the first-order derivative for the auxiliary
excitation energy is presented in
Appendix~\ref{appendix:derivauxXE}. As readily seen from
the expression in Eq.~(\ref{eq:derivauxXE_any_w_expansion_GOK}), at $w=0$, the derivative
contains two terms. The first one, that is linear in the Hxc 
kernel, is expected to be positive due to the Hartree
contribution. The second one is quadratic in the Hxc kernel and
is negative (because of the denominator), exactly like conventional second-order contributions to the
ground-state energy in many-body perturbation theory. The latter term might be large enough at $w=0$
so that the auxiliary excitation energy decreases with increasing $w$.
The linearity in $w$ (last term on the right-hand side of
Eq.~(\ref{eq:derivauxXE_any_w_expansion_GOK})) explains why that derivative
becomes zero and is then positive for larger $w$ values. As the
excitation energy increases, the denominator mentioned previously also
increases. The derivative will therefore increase, thus leading to the
positive curvature observed for the auxiliary
excitation energy. All
these features are essentially driven by the response of the
auxiliary excited state to changes in the ensemble weight (not shown). Returning to
the top panels in
Fig.~\ref{Fig:He_w_curves}, we see that the minimum at $w=0.01$ only
appears when auxiliary energies are computed self-consistently. This is
consistent with Eq.~(\ref{eq:derivauxXE_any_w_expansion_GOK})
where the second (negative) term on the right-hand side describes the
response of the KS orbitals to changes in the Hxc potential through the
$w$-dependent ensemble density. When the latter term is neglected, the auxiliary excitation
energy has positive slope already at $w=0$. For larger $w$ values,
self-consistency effects on the slope are reduced. Indeed, 
the response of the GOK orbitals is expected to be
smaller as the auxiliary excitation energy increases. The large
deviation of the non-self-consistent auxiliary excitation energy from the
self-consistent one is due to the fact that, for the former,
the ensemble density is constructed from the ground-state KS orbitals.
Finally, we note that the self-consistent auxiliary excitation energy
equals the reference FCI one around $w=0.4$. A very similar result has
been obtained at the qLDA level by Yang {\it et al.}~\cite{PRA14_Burke_exact_GOK-DFT} 
We also find that both LDA and PBE yield very similar results.\\ 

Let us now turn to the LIM excitation energy for $\mu=0$. By
construction, it is
$w$-independent, like in the exact theory. Note that the auxiliary
excitation energy equals the LIM one for a $w$ value that is slightly
larger than 1/4, thus showing that the ensemble energy is not strictly
quadratic in $w$. Moreover, as expected from the analysis in
Appendix~\ref{appendix:sc-effects}, the effect of self-consistency is
much stronger on the auxiliary excitation energy than on the LIM one. For
the latter it is actually negligible. Turning to the effective DDs in the top panels of
Fig.~\ref{Fig:He_w_curves}, these qualitatively vary with the ensemble
weight similar to the accurate DD shown in Fig.~7 of
Ref.~\cite{PRA14_Burke_exact_GOK-DFT}. Still, there are significant differences. For $w=0$, the
effective DD equals 0.0736 and 0.0814 $E_h$ at the LDA and PBE levels,
respectively. The accurate value obtained by Yang {\it et
al.}~\cite{PRA14_Burke_exact_GOK-DFT} is much smaller (0.0116 $E_h$). In
addition, both LDA and PBE effective DDs equal zero close to $w=1/4$
that is much smaller than the accurate value of 
Ref.~\cite{PRA14_Burke_exact_GOK-DFT} ($w\approx0.425$). Note finally
that the substantial difference between the LIM and FCI excitation
energies prevents the effective DD and shifted auxiliary excitation energy
curves to be symmetric with respect to the weight axis, as it should be
in the exact theory.\\

\subsubsection{Range-separated results for He}

As illustrated in the middle and bottom panels of Fig.~\ref{Fig:He_w_curves},
the auxiliary excitation energy, shown for $\mu=0.4$ and $1.0a_0^{-1}$, becomes linear in $w$ as $\mu$
increases. This is in agreement with the first-order derivative
expression in Eq.~(\ref{eq:derivauxXE_any_w_expansion}). Indeed, when
$\mu\rightarrow+\infty$, the auxiliary wavefunctions become the physical
ones which are $w$-independent. Consequently, the third term on
the right-hand side, that is responsible for the minimum at $w=0.01$ observed when 
$\mu=0$, vanishes for larger $\mu$ values. Similarly, the auxiliary
energies will become $w$-independent and equal to the physical energies, thus leading to a $w$-independent
first-order derivative. Interestingly, the (negative) second term on the
right-hand side of Eq.~(\ref{eq:derivauxXE_any_w_expansion}) is
quadratic in the short-range kernel and is taken
into account only when calculations are performed self-consistently. 
Since the short-range kernel becomes small as $\mu$ increases, it is not
large enough to compensate the positive contribution from the first term
that is linear in the short-range kernel. As a result, the slope of the
auxiliary excitation energy is
positive for all $w$ values. It also becomes clear that self-consistency
will decrease the slope.\\

Turning to the LIM excitation energies and the effective DDs,
the former become closer to the FCI
value as $\mu$ increases while the latter are reduced, as expected. The fact
that the auxiliary excitation energy equals the LIM one for $w=0.25$
confirms that the range-separated ensemble energy is essentially
quadratic in $w$ when $\mu\geq0.4a_0^{-1}$.    
Even
though no accurate values for the short-range DD are available in the
literature for any $w$, Fig.~2 in Ref.~\cite{RebTouTeaHelSav-JCP-14}
provides reference values for $w=0$ that are about 0.008 and 0.005 $E_h$
for $\mu=0.4$ and $1.0a_0^{-1}$, respectively. These values are simply
obtained by subtracting the auxiliary excitation energies (denoted
$\Delta\mathcal{E}^\mu_k$ in Ref.~\cite{RebTouTeaHelSav-JCP-14}) from
the standard FCI value ($\mu\rightarrow+\infty$ limit). The effective 
DDs computed at the srLDA level for $\mu=0.4$ and $1.0a_0^{-1}$ differ
from these reference values
by about a factor of ten. Note that srLDA and srPBE
functionals give very similar results.\\

\subsubsection{Be and the stretched HeH$^+$ molecule 
}
GOK-LDA and srLDA ($\mu=0.4$ and $1.0a_0^{-1}$) results are
presented for Be and the stretched HeH$^+$ molecule in Fig.~\ref{Fig:Be_and_HeHplus_w_curves}.
In both systems, the
ensemble contains the ground state and a first singly-excited state,
exactly like for He. Effective DD curves share similar patterns but
their interpretations differ substantially. Let us first consider
the Be atom. At the GOK-LDA level (top left-hand panel in
Fig.~\ref{Fig:Be_and_HeHplus_w_curves}), self-consistency effects are
important. They are responsible for the negative slope of the auxiliary
excitation energy at $w=0$. Interestingly, the slope at $w=0$ is larger 
in absolute value for He than for
Be. This is clearly shown in the bottom panel of
Fig.~\ref{Fig:compar_auxXE_HeBe_HeHplus}. As the auxiliary excitation energy decreases on a broader interval
than for He, the second term
on the right-hand side of
Eq.~(\ref{eq:derivauxXE_any_w_expansion_GOK}) might become larger in absolute
value as $w$ increases. Its combination with the third term (linear in
$w$) may explain why the
minimum is reached at a larger ensemble weight value than for He
($w\approx0.045$). One may also argue that this third term, that is only
described at the self-consistent level, is smaller for Be than for He,
thus leading to a less pronounced curvature in $w$, as shown in
the top panel of Fig.~\ref{Fig:compar_auxXE_HeBe_HeHplus}. The auxiliary excitation
energy becomes linear in $w$ when $\mu=0.4$ and $1.0a_0^{-1}$ (see 
middle and bottom left-hand panels in
Fig.~\ref{Fig:Be_and_HeHplus_w_curves}). Note finally that the
effective DDs are about ten times smaller than in He.\\  

Let us now focus on the stretched HeH$^+$ molecule.
As shown in Fig.~\ref{Fig:compar_auxXE_HeBe_HeHplus}, patterns observed
at the GOK-LDA level
for He and Be are strongly enhanced due to the charge transfer. The
interpretation is however quite different. Indeed, as shown in the
top right-hand panel of Fig.~\ref{Fig:Be_and_HeHplus_w_curves},
self-consistency is negligible for small $w$ values and is therefore not
responsible for the large negative slope of the auxiliary excitation
energy at $w=0$. This was expected since the self-consistent
contribution to the slope (second term on the right-hand side of
Eq.~(\ref{eq:derivauxXE_any_w_expansion_GOK})) involves the overlap between
the HOMO (localized on He) and the LUMO which is, in this particular
case, strictly zero. Consequently, as readily seen in
Eq.~(\ref{eq:simplified_kernel_term_LDA_CT}), the (negative)
LDA exchange and correlation kernels~\cite{td-hf-srdft_open_shell_Elisa} are 
responsible for the
negative slope at $w=0$. The latter is actually smaller in absolute value when the
 LDA correlation density functional is set to zero in the calculation (not shown), thus confirming the importance of both exchange and correlation contributions to the kernel. Note that, as $w$ increases, self-consistency
effects are growing. This can be related with the third term on the right-hand side of
Eq.~(\ref{eq:derivauxXE_any_w_expansion_GOK}) where the response of the
excited state to changes in $w$ contributes. Interestingly, for
$\mu=0.4a^{-1}_0$, the contribution to the slope, at $w=0$, from the short-range exchange-correlation kernel is
significant enough~\cite{td-hf-srdft_open_shell_Elisa} so that the
pattern observed at the GOK-LDA level does not completely disappear (see
the middle right-hand panel in Fig.~\ref{Fig:Be_and_HeHplus_w_curves}).
On the other hand, for the larger  $\mu=1.0a^{-1}_0$ value, the
auxiliary excitation energy becomes essentially linear in $w$ with a positive
slope (see the bottom right-hand panel in
Fig.~\ref{Fig:Be_and_HeHplus_w_curves}). Note finally that the stretched HeH$^+$
molecule exhibits the largest effective DDs.

\subsubsection{H$_2$}

Results obtained for H$_2$ are shown in
Figs.~\ref{Fig:compar_auxXE_HeBe_HeHplus} and \ref{Fig:H2_w_curves}. At
equilibrium, they are quite similar to those obtained
for He. Still, at the GOK-LDA level, the negative slope of the auxiliary
excitation energy at $w=0$ is not
related with self-consistency (see the top left-hand panel in
Fig.~\ref{Fig:H2_w_curves}), in contrast to He. Self-consistency effects become significant as $w$
increases. Effective DDs at $w=0$ are equal to 40.9, 36.2 and 8.6 m$E_h$
for $\mu=0$, 0.4 and 1.0$a_0^{-1}$, respectively. They are
significantly larger than the accurate values deduced from Fig.~6 in
Ref.~\cite{RebTouTeaHelSav-JCP-14} (7.1, 5.7 and about
zero m$E_h$).\\

In the stretched geometry (right-hand panels in
Fig.~\ref{Fig:H2_w_curves}), the nature of the first excited state 
completely changes. It corresponds to the double excitation
$1\sigma^2_g\rightarrow1\sigma^2_u$. At the GOK-LDA level,
self-consistency effects are negligible. This was expected since,
according to Eq.~(\ref{eq:derivauxXE_any_w_expansion}), the latter
effects involve couplings between ground and excited states through the
density operator. Consequently, a doubly-excited state will not
contribute. Moreover, the difference in densities between the ground-state and first doubly-excited GOK determinants
reduces along the bond-breaking coordinate, simply because the
overlap between the $1s$ orbitals reduces. As a result,
the first-order derivative of the auxiliary excitation energy is very
small, as confirmed by Fig.~\ref{Fig:compar_auxXE_HeBe_HeHplus}.
This analysis holds also for larger $\mu$ values. The only difference is
that, when $\mu>0$, both ground- and excited-state wavefunctions are
multiconfigurational~\cite{PaolasrXmd,MP15_Manu_soet_cassoft}. In a
minimal basis, they are simply written as 
\begin{eqnarray}
\vert\Psi_0^\mu\rangle&=&\dfrac{1}{\sqrt{2}}\Big(\vert\sigma^2_g\rangle
-
\vert\sigma^2_u\rangle\Big),
\nonumber\\
\vert\Psi_1^\mu\rangle&=&\dfrac{1}{\sqrt{2}}\Big(\vert\sigma^2_g\rangle
+
\vert\sigma^2_u\rangle
\Big).
\end{eqnarray}
In this case, both ground and excited states have the same density,
\begin{eqnarray}
n_{\Psi_0^\mu}(\mathbf{r})=n_{\Psi_1^\mu}(\mathbf{r})=\dfrac{1}{2}\Big(n_{\sigma^2_g}(\mathbf{r})+n_{\sigma^2_u}(\mathbf{r})\Big),
\end{eqnarray}
and their coupling through the
density operator equals 
\begin{eqnarray}\label{eq:coupling_dens_2bleX}
\langle \Psi_0^\mu\vert \hat{n}(\mathbf{r})\vert\Psi_1^\mu\rangle
=\dfrac{1}{2}\Big(n_{\sigma^2_g}(\mathbf{r})-n_{\sigma^2_u}(\mathbf{r})\Big),
\end{eqnarray}  
which is zero as the overlap between the $1s$ orbitals is neglected.\\

Since the ensemble energy is, for any $\mu$ value, almost linear in $w$, the LIM
and auxiliary excitation energies are very close for any weight.
Consequently, the effective DD is very small (4.5 m$E_h$ for 
$\mu=0a_0^{-1}$ and $w=0$). Since the deviation of the LIM excitation
energy from the FCI one is relatively large 
(about $-0.12E_h$ for $\mu=0a_0^{-1}$), symmetry of the plotted curves
with respect to the weight axis is completely broken, in contrast to the
other systems. 
In this
particular situation, LIM brings no improvement and the
effective DD is expected to be far
from the true DD. For comparison, the latter equals about 200 m$E_h$
for a slightly larger bond distance (4.2$a_0$) and $\mu=0a_0^{-1}$,   
according to Fig.~7 in
Ref.~\cite{RebTouTeaHelSav-JCP-14}. For the same distance, the KS-LDA auxiliary
excitation energy (not shown) deviates by 130m$E_h$ in absolute value from the FCI value,
which is in the same order of magnitude as the true DD. Therefore, for
$R=3.7a_0$, the true DD is expected to be much larger than the effective
one.            

\subsection{Excitation energies}

\subsubsection{Single excitations}

LIM excitation energies have been computed when varying $\mu$ for the
various systems studied previously. Single excitations are discussed
in this section. Results are shown in Fig.~\ref{Fig:singles_mu_curves}.
It is quite remarkable that, already for $\mu=0$, LIM performs better
than standard TD-DFT with the same functional (LDA or PBE). This is also
true for the $2\Sigma^+$ charge transfer state in the stretched HeH$^+$
molecule. We even obtain slightly better results than the popular
TD-CAM-B3LYP method. As expected, the error with respect to FCI reduces
as $\mu$ increases. Note that, for He, it becomes zero and then changes
sign in the vicinity
of $\mu=1.0a^{-1}_0$. The latter value gives also accurate results for
the other systems, which is in agreement with
Ref.~\cite{PRA13_Pernal_srEDFT}. Note also that, for the typical value
$\mu= 0.4-0.5a^{-1}_0$~\cite{Angyan2005PRA,JCPunivmu}, the slope in $\mu$ for the LIM excitation
energy is quite significant. It would therefore be relevant to adapt the extrapolation scheme of
Savin~\cite{JCP14_Andreas_extrapol_range_sep,PRA15_Elisa_extrapolation_XE_AC}
to range-separated ensemble DFT. This is left for future work. Note that
srLDA and srPBE functionals give rather similar results. For comparison, auxiliary 
excitation energies obtained from the ground-state density ($w=0$) are also shown. 
The former reduce to KS orbital energy differences for $\mu=0$. In
this case, TD-DFT gives slightly better results, except for the charge
transfer excitation in HeH$^+$ where the difference is very small, as
expected~\cite{Casida_tddft_review_2012}. Both srLDA and srPBE auxiliary excitation
energies reach a minimum at relatively small $\mu$ values
(0.125$a^{-1}_0$ for He). This is due to the approximate short-range
(semi-)local potentials that we used. Indeed, as shown in
Ref.~\cite{RebTouTeaHelSav-JCP-14}, variations in $\mu$ are expected to be monotonic for He and H$_2$
at equilibrium if an accurate
short-range potential were used. Since the range-separated ensemble
energy can be expressed in terms of the auxiliary energies (see
Eq.~(\ref{eq:RS_Eens_with_auxE})), it is not surprising to recover such 
minima for some LIM excitation energies. Let us finally note that
the auxiliary excitation energy converges more rapidly than the LIM one to the FCI
value when $\mu$ increases from 1.0$a_0^{-1}$. For Be, convergences are
very similar. As already mentioned, the convergence can actually be
further improved by means of 
extrapolation
techniques~\cite{JCP14_Andreas_extrapol_range_sep,PRA15_Elisa_extrapolation_XE_AC}.      
In conclusion, the LIM approach is promising at both GOK-DFT and
range-separated ensemble DFT levels. In the latter case, $\mu$ should
not be too large otherwise the use of an ensemble is less relevant.
Indeed, auxiliary excitation energies
obtained from the ground-state density are in fact better approximations
to the FCI excitation energies, at least for the systems and
approximate short-range functionals considered in this work. This should
be tested on more systems in the future.

\subsubsection{Double excitations}

One important feature of both GOK and range-separated ensemble DFT is
the possibility of modeling multiple excitations, in contrast to
standard TD-DFT. Results obtained for the $2^1\Sigma_g^+$ and $1^1D$
states in the stretched H$_2$ molecule and Be, respectively, are shown
in Fig.~\ref{Fig:LIM_2ble_XE}. We focus on H$_2$ first. As discussed
previously, LIM and auxiliary excitation energies are almost identical in this
case. For $\mu=0a_0^{-1}$, they differ by about -0.12 $E_h$ from the FCI
value. There are no significant differences between srLDA and srPBE
results.      
The error monotonically reduces with increasing $\mu$. Interestingly,
for $\mu=0.4a_0^{-1}$, the LIM excitation energy equals 0.237$E_h$, that is
very similar to the multi-configuration range-separated TD-DFT result
obtained with the same functionals (0.238$E_h$).~\cite{fromager2013} This confirms that the
short-range kernel does not contribute significantly to the excitation energy, since
the ground and doubly-excited states are not coupled by the density
operator (see Eq.~(\ref{eq:coupling_dens_2bleX})). Note that, for
$R=4.2a_0$ and $\mu=0.4a_0^{-1}$, the srLDA auxiliary excitation energy
(not shown) equals $0.194E_h$, that is rather close to the accurate value
($0.181E_h$) deduced from Fig.~7 in Ref.~\cite{RebTouTeaHelSav-JCP-14}.
As a result, the approximate (semi-)local density-functional potentials are not responsible
for the large error on the excitation energy. One would blame the adiabatic approximation if TD
linear response theory were used. In our case, it is related to the WIDFA
approach. In this respect, it seems essential to develop
weight-dependent exchange-correlation functionals for ensembles. 
Applying the 
GACE formalism to model
systems would be instructive in that respect. Work
is currently in progress in this direction.\\          

Turning to the doubly-excited $1^1D$ state in Be, LIM is quite accurate
already at the GOK-DFT level. Interestingly, the largest and relatively
small errors in
absolute value (about 4.0 and 7.0 m$E_h$ for the srLDA and srPBE
functionals, respectively) are obtained around $\mu=1.0a_0^{-1}$. In this case,
the ensemble contains four states ($1^1S$, $2^1S$ and two degenerate
$1^1D$ states) whereas in all previous cases first excitation energies
were computed with only two states. This indicates that $\mu$ values
that are optimal in terms of accuracy may depend on the choice of the
ensemble. This should be investigated further in the future.  


\section{Conclusions}\label{sec:conclusions}

A rigorous combination of wavefunction theory with ensemble DFT for
excited states has been investigated by means of range separation. As
illustrated for simple two- and four-electron systems, using local or semi-local ground-state density-functional approximations for modeling the short-range exchange-correlation energy of a  
bi-ensemble with weight $w$ usually leads to range-separated ensemble energies that are not
strictly linear in $w$. Consequently, the approximate excitation energy,
that is defined as the derivative of the ensemble energy with respect to
$w$, becomes $w$-dependent, unlike the exact derivative. Moreover,   
the variation in $w$ can be very sensitive to the self-consistency
effects that are induced by the short-range density-functional potential.\\

 In order to
define unambiguously approximate excitation energies in this context, we
proposed a linear interpolation method (LIM) that simply interpolates
the ensemble energy between $w=0$ (ground state) and $w=1/2$
(equiensemble consisting of the non-degenerate ground and
first excited states). A generalization to higher excitations with
degenerate ground and excited states has been formulated and tested. It simply consists in
interpolating the ensemble energy linearly between equiensembles.
LIM is applicable to GOK-DFT that is
recovered when the range-separation parameter $\mu$ equals zero. In the
latter case, LIM performs systematically better 
than standard TD-DFT with the same functional, even for the
$2\Sigma^+$ charge-transfer state in the stretched HeH$^+$ molecule. For typical values
$\mu=0.4-0.5a_0^{-1}$, LIM gives a better approximation to the
excitation energy than the auxiliary long-range-interacting one obtained
from the ground-state density. However, for larger $\mu$ values, the
latter excitation energy usually converges faster than the LIM one to
the physical result.\\

 One of the motivation for using ensembles is the
possibility, in contrast to standard TD-DFT, to model
double excitations. Results obtained with LIM for the $1^1D$ state in Be are
relatively accurate, especially at the GOK-DFT level. In the particular
case of the stretched H$_2$ molecule, the range-separated ensemble
energy is almost linear in $w$, thus making the approximate
$2^1\Sigma_g^+$ excitation energy almost weight-independent. LIM brings
no improvement in that case and the error on the excitation energy is
quite significant. This example illustrates the need for
weight-dependent exchange-correlation functionals. Combining adiabatic
connection formalisms~\cite{MP14_Manu_GACE} with accurate reference
data~\cite{PRA14_Burke_exact_GOK-DFT} will hopefully enable
the development of density-functional approximations for ensembles in the
near future.\\
 
Finally, in order to turn LIM into a useful modelling tool,
a state-averaged complete active space self-consistent field
(SA-CASSCF) should be used rather than CI for the computation of
long-range correlation effects. Since the long-range interaction has
no singularity at $r_{12}=0$, we expect a limited
number of configurations to be sufficient for recovering most of the
long-range correlation. This observation has already been made for the
ground state~\cite{nevpt2srdft,JCP15_Odile_basis_convergence_srDFT}.
Obviously, the active space should be chosen carefully in order to
preserve size consistency. The implementation and calibration of such a
method is left for future work.  

\begin{acknowledgments}
E.F. thanks Alex Borgoo and Laurent Mazouin for fruitful discussions. The authors would like
to dedicate the paper to the memory of Prof.~Tom Ziegler who supported
this work on ensemble DFT and contributed
significantly in recent years to the development of time-independent DFT for excited
states. E.F.~acknowledges financial support from
LABEX "Chemistry of complex systems"
and ANR (MCFUNEX project).   
\end{acknowledgments}

\appendix
    \setcounter{equation}{0}  
 
      \section{SELF-CONSISTENT RANGE-SEPARATED ENSEMBLE
DENSITY-FUNCTIONAL PERTURBATION THEORY
	}\label{appendix:sc_pt}  

The self-consistent Eq.~(\ref{eq:sc_widfa_eq}) can be solved for small
$w$ values within
perturbation theory. For that purpose we partition the long-range interacting
density-functional Hamiltonian as follows,
\begin{eqnarray}
\hat{H}^\mu[\tilde{n}^{\mu,w}]=\hat{H}^\mu[{n}^0]+w\hat{\mathcal{W}}^{\mu,w},
\end{eqnarray}
where, according to Eq.~(\ref{eq:lrhamil}), the perturbation equals
\begin{eqnarray}\label{eq:sc-pt_perturbation}
\hspace{-0.4cm}w\hat{\mathcal{W}}^{\mu,w}=
\int \ddroit\mathbf{r}
\Bigg(
\dfrac{\delta E^{{\rm sr,\mu}}_{\rm
Hxc}[\tilde{n}^{\mu,w}]}{\delta
n(\mathbf{r})}
-\dfrac{\delta E^{{\rm sr,\mu}}_{\rm
Hxc}[n^0]}{\delta
n(\mathbf{r})}
\Bigg)
\hat{n}(\mathbf{r}),
\end{eqnarray}
and, according to Eq.~(\ref{eq:aux_density_widfa}),
\begin{eqnarray}\label{eq:Taylor_exp_Edens_1storder}
\tilde{n}^{\mu,w}(\mathbf{r})&=&
n^0(\mathbf{r})+\left. w
\dfrac{\partial
\tilde{n}^{\mu,w}(\mathbf{r})}{\partial w}
\right |_{w=0}+\mathcal{O}(w^2)
\nonumber\\
&=&
n^0(\mathbf{r})+
w\Big(n_{\Psi{^\mu_1}}(\mathbf{r})-n^0(\mathbf{r})\Big)
\nonumber\\
&&
+
\left. w
\dfrac{\partial
{n}_{\tilde{\Psi}{^{\mu,w}_0}}(\mathbf{r})}{\partial w}
\right |_{w=0}
+\mathcal{O}(w^2)
.
\end{eqnarray}
Combining Eq.~(\ref{eq:sc-pt_perturbation}) with
Eq.~(\ref{eq:Taylor_exp_Edens_1storder}) leads to
\begin{eqnarray}\label{eq:sc-pt_perturbation_1storder}
\hat{\mathcal{W}}^{\mu,w}&=&
\hat{\mathcal{W}}^{\mu,0}+\mathcal{O}(w)
\nonumber\\
&=&
\int \int\ddroit\mathbf{r}\ddroit\mathbf{r'}\dfrac{\delta^2E^{{\rm
sr,\mu}}_{\rm Hxc}[n^0]}{\delta n(\mathbf{r'})\delta n(\mathbf{r})}
\bigg(
n_{\Psi{^\mu_1}}(\mathbf{r'})-n^0(\mathbf{r'})
\nonumber\\
&&+\left.\dfrac{\partial
n_{\tilde{\Psi}{^{\mu,w}_0}}(\mathbf{r'})}{\partial w}\right|_{w=0}
\bigg)
\hat{n}(\mathbf{r})+\mathcal{O}(w).
\end{eqnarray}
From the usual first-order wavefunction correction expression
\begin{eqnarray}
\left.\Big\vert\dfrac{\partial \tilde{\Psi}{^{\mu,w}_0}}{\partial w}
\Big\rangle
\right|_{w=0}
=\sum_{i\geq 1}\vert\Psi^\mu_i\rangle\dfrac{\langle\Psi^\mu_i \vert
\hat{\mathcal{W}}^{\mu,0}\vert\Psi^\mu_0\rangle}{\mathcal{E}_0^\mu-\mathcal{E}_i^\mu},
\end{eqnarray}
and the expression for the derivative of the ground-state density, that
we simply denote $\partial n^{\mu}$,
\begin{eqnarray}\label{eq:partial_n_def}
\partial n^{\mu}(\mathbf{r_1})&=&\left.\dfrac{\partial
n_{\tilde{\Psi}{^{\mu,w}_0}}(\mathbf{r_1})}{\partial w}\right|_{w=0}
\nonumber\\
&=&2\left.\Big\langle\Psi^\mu_0\Big\vert 
\hat{n}(\mathbf{r_1})\Big \vert\dfrac{\partial \tilde{\Psi}{^{\mu,w}_0}}{\partial w}
\Big\rangle
\right|_{w=0},
\end{eqnarray}
we obtain the self-consistent equation
\begin{eqnarray}\label{eq:sc_for_partial_n_w0}
\partial n^{\mu}=\hat{\mathcal{F}}\partial
n^{\mu}+\hat{\mathcal{F}}\big(n_{\Psi{^\mu_1}}-n^0\big),
\end{eqnarray}
where $\hat{\mathcal{F}}$ is a linear operator that acts on any function
$f(\mathbf{r})$ as follows,
\begin{eqnarray}\label{eq:Foperator}
&&\hat{\mathcal{F}}f(\mathbf{r_1})
=
2\sum_{i\geq 1}
\int \int\ddroit\mathbf{r}\ddroit\mathbf{r'}
\dfrac{\delta^2E^{{\rm
sr,\mu}}_{\rm Hxc}[n^0]}{\delta n(\mathbf{r'})\delta n(\mathbf{r})}
\dfrac{n_{0i}^\mu(\mathbf{r_1})n^\mu_{0i}(\mathbf{r})}{\mathcal{E}_0^\mu-\mathcal{E}_i^\mu}f(\mathbf{r'})
,
\nonumber\\
&&{n}^\mu_{0i}(\mathbf{r})=\langle \Psi^\mu_0\vert
\hat{n}(\mathbf{r})\vert\Psi^\mu_i\rangle. 
\end{eqnarray}
Consequently,
\begin{eqnarray}\label{eq:sc_rsp_GS_density}
\partial
n^{\mu}&=&\big(1-\hat{\mathcal{F}}\big)^{-1}\hat{\mathcal{F}}\big(n_{\Psi{^\mu_1}}-n^0\big)
\nonumber\\
&=&\sum^{+\infty}_{k=0}\hat{\mathcal{F}}^k\hat{\mathcal{F}}\big(n_{\Psi{^\mu_1}}-n^0\big)
\nonumber\\
&=&\hat{\mathcal{F}}\big(n_{\Psi{^\mu_1}}-n^0\big)+\ldots
\end{eqnarray}

      \section{DERIVATIVE OF THE AUXILIARY EXCITATION ENERGY
	}\label{appendix:derivauxXE}  

According to Eq.~(\ref{eq:hell-Feyn_widfa}), the first-order derivative of the individual auxiliary energies can be expressed as 
\begin{eqnarray}
\dfrac{\ddroit \tilde{\mathcal{E}}^{\mu,w}_i}{\ddroit w}=
\int \int\ddroit\mathbf{r'}\ddroit\mathbf{r}&&\dfrac{\delta^2E^{{\rm
sr,\mu}}_{\rm Hxc}[\tilde{n}^{\mu,w}]}{\delta n(\mathbf{r'})\delta n(\mathbf{r})}
\nonumber\\ 
&&
\times\dfrac{\partial
\tilde{n}^{\mu,w}(\mathbf{r}')}{\partial w}n_{\tilde{\Psi}^{\mu,w}_i}(\mathbf{r}),
\end{eqnarray}
where
\begin{eqnarray}\label{eq:rsp_ensemble_density_any_w}
\dfrac{\partial
\tilde{n}^{\mu,w}(\mathbf{r}')}{\partial w}
&=&
\delta \tilde{n}^{\mu,w}(\mathbf{r'})
+
\dfrac{\partial
{n}_{\tilde{\Psi}{^{\mu,w}_0}}(\mathbf{r'})}{\partial w}
\nonumber\\
&&
+w
\dfrac{\partial
\delta \tilde{n}^{\mu,w}(\mathbf{r'})}{\partial w},
\end{eqnarray}
and
\begin{eqnarray}
\delta \tilde{n}^{\mu,w}(\mathbf{r'})=n_{\tilde{\Psi}^{\mu,w}_1}(\mathbf{r'})-n_{\tilde{\Psi}^{\mu,w}_0}(\mathbf{r'}),
\end{eqnarray}
so that the derivative of the auxiliary excitation energy in Eq.~(\ref{eq:exactderivative_tilde_RS_energy}) can be written as
\begin{eqnarray}\label{eq:derivauxXE_any_w}
&&\dfrac{\ddroit \Delta\tilde{\mathcal{E}}^{\mu,w}}{\ddroit w}=
\int \int\ddroit\mathbf{r'}\ddroit\mathbf{r}\dfrac{\delta^2E^{{\rm
sr,\mu}}_{\rm Hxc}[\tilde{n}^{\mu,w}]}{\delta n(\mathbf{r'})\delta n(\mathbf{r})}
\nonumber\\ 
&&
\times \Bigg(\delta \tilde{n}^{\mu,w}(\mathbf{r'})\delta \tilde{n}^{\mu,w}(\mathbf{r})
+
\dfrac{\partial
{n}_{\tilde{\Psi}{^{\mu,w}_0}}(\mathbf{r'})}{\partial w}\delta \tilde{n}^{\mu,w}(\mathbf{r})
\nonumber\\
&&+w
\dfrac{\partial
\delta \tilde{n}^{\mu,w}(\mathbf{r'})}{\partial w}\delta \tilde{n}^{\mu,w}(\mathbf{r})
\Bigg).
\end{eqnarray}
According to perturbation theory through first order (see
Appendix~\ref{appendix:sc_pt}), the response of
the ground-state density to variations in the ensemble weight equals   
\begin{eqnarray}\label{eq:variations_n_0_w}
 \dfrac{\partial
n_{\tilde{\Psi}{^{\mu,w}_0}}(\mathbf{r'})}{\partial w}
&=&2\Big\langle\tilde{\Psi}^{\mu,w}_0\Big\vert 
\hat{n}(\mathbf{r'})\Big \vert\dfrac{\partial \tilde{\Psi}{^{\mu,w}_0}}{\partial w}
\Big\rangle
\nonumber\\
&=&2
\sum_{i\geq 1}
\int \int\ddroit\mathbf{r_1}\ddroit\mathbf{r_2}
\dfrac{\delta^2E^{{\rm
sr,\mu}}_{\rm Hxc}[\tilde{n}^{\mu,w}]}{\delta n(\mathbf{r_2})\delta n(\mathbf{r_1})}
\nonumber\\
&&\times
\dfrac{n_{0i}^{\mu,w}(\mathbf{r'})n^{\mu,w}_{0i}(\mathbf{r_1})}{\tilde{\mathcal{E}}_0^{\mu,w}-\tilde{\mathcal{E}}_i^{\mu,w}}
\dfrac{\partial
\tilde{n}^{\mu,w}(\mathbf{r_2})}{\partial w}
,
\end{eqnarray}
where
$n_{0i}^{\mu,w}(\mathbf{r'})=\langle\tilde{\Psi}{^{\mu,w}_0}\vert\hat{n}(\mathbf{r'})\vert\tilde{\Psi}{^{\mu,w}_i}\rangle$.
Note that, as already pointed out for $w=0$ (see
Eq.~(\ref{eq:sc_for_partial_n_w0})), Eq.~(\ref{eq:variations_n_0_w})
should be solved self-consistently. By considering the first
contribution to the response of the ensemble density in
Eq.~(\ref{eq:rsp_ensemble_density_any_w}) we
obtain
\begin{eqnarray}\label{eq:variations_n_0_w_approx}
&& \dfrac{\partial
n_{\tilde{\Psi}{^{\mu,w}_0}}(\mathbf{r'})}{\partial w}
=2
\sum_{i\geq 1}
\int \int\ddroit\mathbf{r_1}\ddroit\mathbf{r_2}
\dfrac{\delta^2E^{{\rm
sr,\mu}}_{\rm Hxc}[\tilde{n}^{\mu,w}]}{\delta n(\mathbf{r_2})\delta n(\mathbf{r_1})}
\nonumber\\
&&\times
\dfrac{n_{0i}^{\mu,w}(\mathbf{r'})n^{\mu,w}_{0i}(\mathbf{r_1})}{\tilde{\mathcal{E}}_0^{\mu,w}-\tilde{\mathcal{E}}_i^{\mu,w}}
\delta \tilde{n}^{\mu,w}(\mathbf{r_2})+\ldots
\end{eqnarray}
thus leading to the following expansion 
\begin{eqnarray}\label{eq:derivauxXE_any_w_expansion}
&&\dfrac{\ddroit \Delta\tilde{\mathcal{E}}^{\mu,w}}{\ddroit w}=
\int \int\ddroit\mathbf{r'}\ddroit\mathbf{r}\dfrac{\delta^2E^{{\rm
sr,\mu}}_{\rm Hxc}[\tilde{n}^{\mu,w}]}{\delta n(\mathbf{r'})\delta n(\mathbf{r})}
\delta \tilde{n}^{\mu,w}(\mathbf{r'})\delta \tilde{n}^{\mu,w}(\mathbf{r})
\nonumber
\\
&&
+2\sum_{i\geq
1}
\dfrac{1}{\tilde{\mathcal{E}}_0^{\mu,w}-\tilde{\mathcal{E}}_i^{\mu,w}}
\nonumber\\
&&\times\Bigg(
\int \int\ddroit\mathbf{r'}\ddroit\mathbf{r}\dfrac{\delta^2E^{{\rm
sr,\mu}}_{\rm Hxc}[\tilde{n}^{\mu,w}]}{\delta n(\mathbf{r'})\delta n(\mathbf{r})}
\delta \tilde{n}^{\mu,w}(\mathbf{r})n_{0i}^{\mu,w}(\mathbf{r'}) 
\Bigg)^2
\nonumber
\\
&&+w
\Bigg(\int \int\ddroit\mathbf{r'}\ddroit\mathbf{r}\dfrac{\delta^2E^{{\rm
sr,\mu}}_{\rm Hxc}[\tilde{n}^{\mu,w}]}{\delta n(\mathbf{r'})\delta n(\mathbf{r})}
\dfrac{\partial
\delta \tilde{n}^{\mu,w}(\mathbf{r'})}{\partial w}\delta \tilde{n}^{\mu,w}(\mathbf{r})
\Bigg)
\nonumber\\
&&+\ldots
\end{eqnarray}
Note that, at the srLDA level of approximation, the exchange-correlation contribution to the short-range kernel is strictly local~\cite{td-hf-srdft_open_shell_Elisa}. By using the decomposition
\begin{eqnarray}
\dfrac{\delta^2E^{{\rm srLDA,\mu}}_{\rm Hxc}[n]}{\delta n(\mathbf{r'})\delta n(\mathbf{r})}
&=&
w^{\rm sr,\mu}_{\rm ee}(\vert\mathbf{r}-\mathbf{r'}\vert)
\nonumber\\
&&+
\dfrac{\partial^2e^{{\rm sr,\mu}}_{\rm xc}(n(\mathbf{r}))}{\partial n^2}\delta(\mathbf{r}-\mathbf{r'}),
\end{eqnarray}
the first term on the right-hand side of Eq.~(\ref{eq:derivauxXE_any_w_expansion}) can be simplified as follows, 
\begin{eqnarray}\label{eq:simplified_kernel_term_srLDA}
&&
\int \int\ddroit\mathbf{r'}\ddroit\mathbf{r}\dfrac{\delta^2E^{{\rm
srLDA,\mu}}_{\rm Hxc}[\tilde{n}^{\mu,w}]}{\delta n(\mathbf{r'})\delta n(\mathbf{r})}
\delta \tilde{n}^{\mu,w}(\mathbf{r'})\delta \tilde{n}^{\mu,w}(\mathbf{r})
\nonumber\\
&&=
\int \int\ddroit\mathbf{r'}\ddroit\mathbf{r}
\,w^{\rm sr,\mu}_{\rm ee}(\vert\mathbf{r}-\mathbf{r'}\vert)\delta \tilde{n}^{\mu,w}(\mathbf{r'})\delta \tilde{n}^{\mu,w}(\mathbf{r})
\nonumber\\
&&+\int\ddroit\mathbf{r}\dfrac{\partial^2e^{{\rm sr,\mu}}_{\rm xc}(\tilde{n}^{\mu,w}(\mathbf{r}))}{\partial n^2}
\Big(\delta \tilde{n}^{\mu,w}(\mathbf{r})\Big)^2.
\end{eqnarray}

In the GOK-DFT limit ($\mu=0$), if the first excitation is a single
excitation from the HOMO to the LUMO, the auxiliary excitation energy
reduces to an orbital energy difference $\Delta\tilde{\varepsilon}^w$
whose derivative can formally be expressed as follows, according to Eq.~(\ref{eq:derivauxXE_any_w_expansion}), 
\begin{eqnarray}\label{eq:derivauxXE_any_w_expansion_GOK}
&&\dfrac{\ddroit \Delta\tilde{\varepsilon}^{w}}{\ddroit w}=
\int \int\ddroit\mathbf{r'}\ddroit\mathbf{r}\dfrac{\delta^2E
_{\rm Hxc}[\tilde{n}^{w}]}{\delta n(\mathbf{r'})\delta n(\mathbf{r})}
\delta \tilde{n}^{w}(\mathbf{r'})\delta \tilde{n}^{w}(\mathbf{r})
\nonumber
\\
&&
+4\sum_{i\leq
N/2,a>N/2}
\dfrac{1}{\tilde{\varepsilon}_i^{w}-\tilde{\varepsilon}_a^{w}}
\nonumber\\
&&\times\Bigg(
\int \int\ddroit\mathbf{r'}\ddroit\mathbf{r}\dfrac{\delta^2E
_{\rm Hxc}[\tilde{n}^{w}]}{\delta n(\mathbf{r'})\delta n(\mathbf{r})}
\delta
\tilde{n}^{w}(\mathbf{r})\tilde{\phi}^w_i(\mathbf{r'})\tilde{\phi}^w_a(\mathbf{r'}) 
\Bigg)^2
\nonumber
\\
&&+w
\Bigg(\int \int\ddroit\mathbf{r'}\ddroit\mathbf{r}\dfrac{\delta^2E
_{\rm Hxc}[\tilde{n}^{w}]}{\delta n(\mathbf{r'})\delta n(\mathbf{r})}
\dfrac{\partial
\delta \tilde{n}^{w}(\mathbf{r'})}{\partial w}\delta \tilde{n}^{w}(\mathbf{r})
\Bigg)
\nonumber\\
&&+\ldots
\end{eqnarray}
where
\begin{eqnarray}
\tilde{n}^{w}(\mathbf{r})&=&2\sum^{N/2-1}_{k=1}\tilde{\phi}^w_k(\mathbf{r})^2
\nonumber\\
&&+(2-w)\tilde{\phi}^w_{N/2}(\mathbf{r})^2+w\tilde{\phi}^w_{N/2+1}(\mathbf{r})^2
,
\nonumber\\
\delta
\tilde{n}^{w}(\mathbf{r})&=&\tilde{\phi}^w_{N/2+1}(\mathbf{r})^2-\tilde{\phi}^w_{N/2}(\mathbf{r})^2,
\end{eqnarray}
and $\{\tilde{\phi}^w_k(\mathbf{r})\}_k$ are the GOK-DFT orbitals
with the associated energies $\{\tilde{\varepsilon}^w_k\}_k$ that
are obtained within the WIDFA approximation. Note that, in practical
calculations, partially occupied GOK-DFT orbitals have not been computed
explicitly. Instead, we performed FCI calculations in the basis of
determinants constructed from the KS orbitals.\\
Let us finally note that if the HOMO and LUMO do not overlap, the first term on the right-hand side of Eq.~(\ref{eq:derivauxXE_any_w_expansion_GOK}) can be further simplified at the LDA level, according to Eq.~(\ref{eq:simplified_kernel_term_srLDA}), thus leading to
\begin{eqnarray}\label{eq:simplified_kernel_term_LDA_CT}
&&
\int \int\ddroit\mathbf{r'}\ddroit\mathbf{r}\dfrac{\delta^2E^{{\rm
LDA}}_{\rm Hxc}[\tilde{n}^{w}]}{\delta n(\mathbf{r'})\delta n(\mathbf{r})}
\delta \tilde{n}^{w}(\mathbf{r'})\delta \tilde{n}^{w}(\mathbf{r})
\nonumber\\
&&\rightarrow
\int \int\ddroit\mathbf{r'}\ddroit\mathbf{r}
\,\dfrac{\tilde{\phi}^w_{N/2}(\mathbf{r})^2\tilde{\phi}^w_{N/2}(\mathbf{r'})^2}{\vert\mathbf{r}-\mathbf{r'}\vert}
\nonumber\\
&&+
\int \int\ddroit\mathbf{r'}\ddroit\mathbf{r}
\,\dfrac{\tilde{\phi}^w_{N/2+1}(\mathbf{r})^2\tilde{\phi}^w_{N/2+1}(\mathbf{r'})^2}{\vert\mathbf{r}-\mathbf{r'}\vert}
\nonumber\\
&&+\int\ddroit\mathbf{r}\dfrac{\partial^2e_{\rm xc}(\tilde{n}^{w}(\mathbf{r}))}{\partial n^2}
\nonumber\\
&&\hspace{1.2cm}\times\Big(\tilde{\phi}^w_{N/2}(\mathbf{r})^4+\tilde{\phi}^w_{N/2+1}(\mathbf{r})^4\Big).
\end{eqnarray}


      \section{SELF-CONSISTENCY EFFECTS ON THE ENSEMBLE AND AUXILIARY
ENERGIES
	}\label{appendix:sc-effects}  

Let $n$ denote a trial ensemble density for which the auxiliary
wavefunctions can be determined: 
\begin{eqnarray}\label{eq:aux_ens_dens}
\hat{H}^{\mu}[n]\vert\Psi^\mu_i[n]\rangle=\mathcal{E}^\mu_i[n]\vert\Psi^\mu_i[n]\rangle,
\hspace{0.2cm} i=0,1.
\end{eqnarray}
The resulting auxiliary ensemble density,
\begin{eqnarray}
n^{w}[n](\mathbf{r})=(1-w)n_{\Psi^\mu_0[n]}(\mathbf{r})+wn_{\Psi^\mu_1[n]}(\mathbf{r}),
\end{eqnarray}
is then a functional of $n$, like the ensemble energy that can be
expressed as
\begin{eqnarray}
&&E^{\mu,w}[n]=(1-w)\mathcal{E}^\mu_0[n]+w\mathcal{E}^\mu_1[n]
\nonumber\\
&&-
\int\ddroit\mathbf{r}\,\dfrac{\delta E_{\rm Hxc}^{\rm
sr,\mu}[n]}{\delta n(\mathbf{r})}n^{w}[n](\mathbf{r})
+E_{\rm Hxc}^{\rm
sr,\mu}[n^{w}[n]].
\end{eqnarray}    
The converged ensemble density $\tilde n^{\mu,w}$ fulfils the following
condition:
\begin{eqnarray}\label{eq:converged_dens_cond}
n^w[\tilde n^{\mu,w}]=\tilde n^{\mu,w}.
\end{eqnarray}
If we now consider variations around the trial density, $n\rightarrow
n+\delta n$, the ensemble energy will vary through first order in
$\delta n$ as follows,
\begin{eqnarray}\label{eq:delta_Eens_trial}
&&\delta E^{\mu,w}[n]=(1-w)\delta \mathcal{E}^\mu_0[n]+w\delta\mathcal{E}^\mu_1[n]
\nonumber\\
&&-\int\ddroit\mathbf{r}\,\delta\Bigg(\dfrac{\delta E_{\rm Hxc}^{\rm
sr,\mu}[n]}{\delta n(\mathbf{r})}n^{w}[n](\mathbf{r})\Bigg)
\nonumber\\
&&+
\int\ddroit\mathbf{r}\,\dfrac{\delta E_{\rm Hxc}^{\rm
sr,\mu}[n^{w}[n]]}{\delta n(\mathbf{r})}\delta n^{w}[n](\mathbf{r})
,
\end{eqnarray}    
where, according to the Hellmann--Feynman theorem,
\begin{eqnarray}\label{eq:delta_aux_ener}
\delta
\mathcal{E}^\mu_i[n]=
\int\ddroit\mathbf{r}\,\delta\Bigg(\dfrac{\delta
E_{\rm Hxc}^{\rm sr,\mu}[n]}{\delta
n(\mathbf{r})}\Bigg)n_{\Psi^\mu_i[n]}(\mathbf{r}).
\end{eqnarray}
Combining Eqs.~(\ref{eq:aux_ens_dens}) and (\ref{eq:delta_Eens_trial})
with Eq.~(\ref{eq:delta_aux_ener}) leads to
\begin{eqnarray}\label{eq:delta_Eens_trial_bis}
\delta E^{\mu,w}[n]=
\int\ddroit\mathbf{r}\,
\Bigg(
&&
\dfrac{\delta E_{\rm Hxc}^{\rm
sr,\mu}[n^{w}[n]]}{\delta n(\mathbf{r})}
-
\dfrac{\delta E_{\rm Hxc}^{\rm
sr,\mu}[n]}{\delta n(\mathbf{r})}
\Bigg)
\nonumber\\
&&
\times\delta n^{w}[n](\mathbf{r})
.
\end{eqnarray}    
According to Eq.~(\ref{eq:delta_aux_ener}), the auxiliary excitation
energy $\Delta\mathcal{E}^\mu[n]=\mathcal{E}^\mu_1[n]-\mathcal{E}^\mu_0[n]$ will vary through first order as
\begin{eqnarray}\label{eq:delta_auxXE_trial}
\delta\Delta\mathcal{E}^\mu[n]=
\int\int
&&\ddroit\mathbf{r}\ddroit\mathbf{r'}\,
\dfrac{\delta^2
E_{\rm Hxc}^{\rm sr,\mu}[n]}{
\delta
n(\mathbf{r'})
\delta
n(\mathbf{r})
}\delta n(\mathbf{r'})
\nonumber\\
&&\times\Big(
n_{\Psi^\mu_1[n]}(\mathbf{r})
-
n_{\Psi^\mu_0[n]}(\mathbf{r})
\Big).
\end{eqnarray}
We conclude from Eqs.~(\ref{eq:converged_dens_cond}),
(\ref{eq:delta_Eens_trial_bis}) and
(\ref{eq:delta_auxXE_trial}) that variations $\delta n$ around the converged ensemble density
$\tilde n^{\mu,w}$ will induce at least first and second order
deviations in $\delta n$ for the auxiliary
excitation and ensemble energies, respectively.



\providecommand{\Aa}[0]{Aa}
%

\clearpage

\textbf{FIGURE CAPTIONS}

\begin{figure}
\caption{\label{Fig:curvature_in_He} Senjean et al, Phys. Rev. A}
\vspace{1cm}
\begin{center}
\begin{tabular}{c}
\resizebox{15cm}{!}{\includegraphics{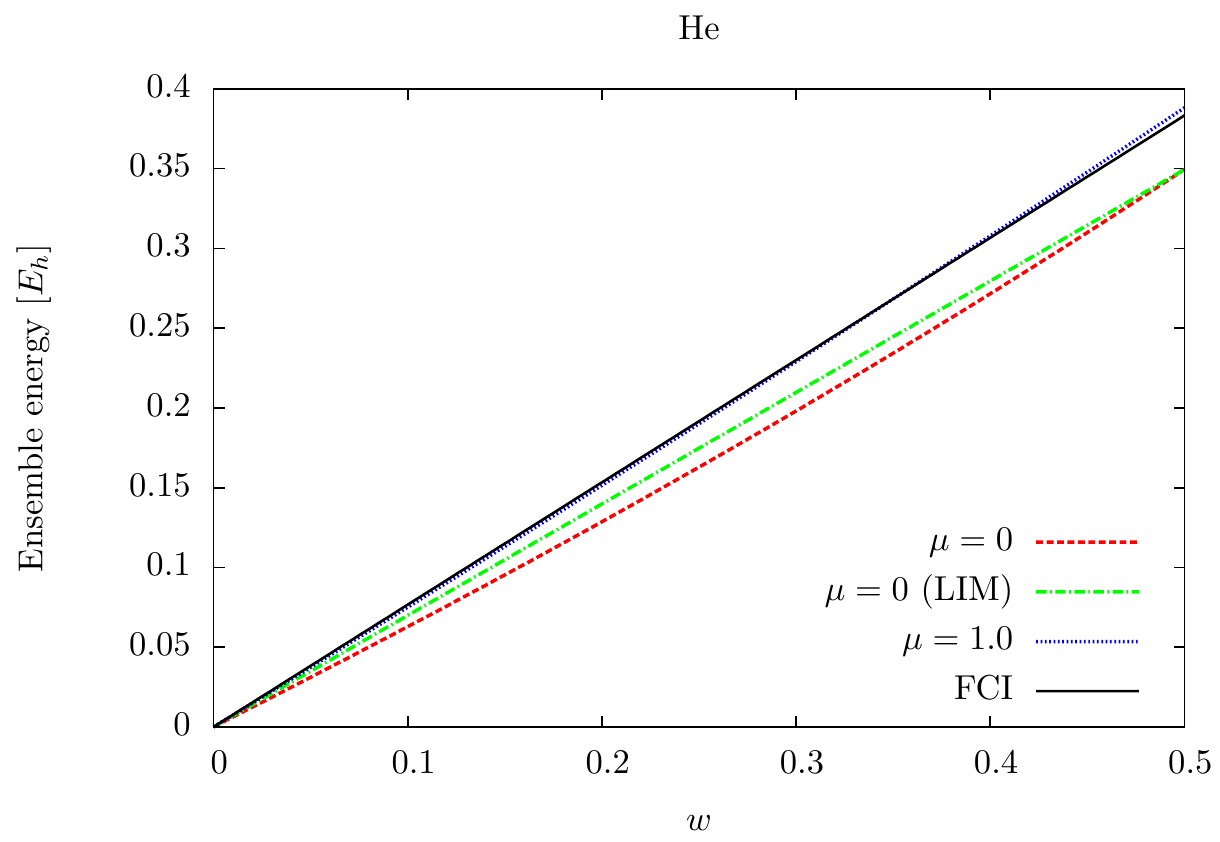}}
\end{tabular}
\end{center}
\end{figure}

\begin{figure}
\caption{\label{Fig:curvature_corr_illustration} Senjean et al, Phys. Rev. A}
\vspace{1cm}
\begin{center}
\begin{tabular}{c}
\resizebox{15cm}{!}{\includegraphics{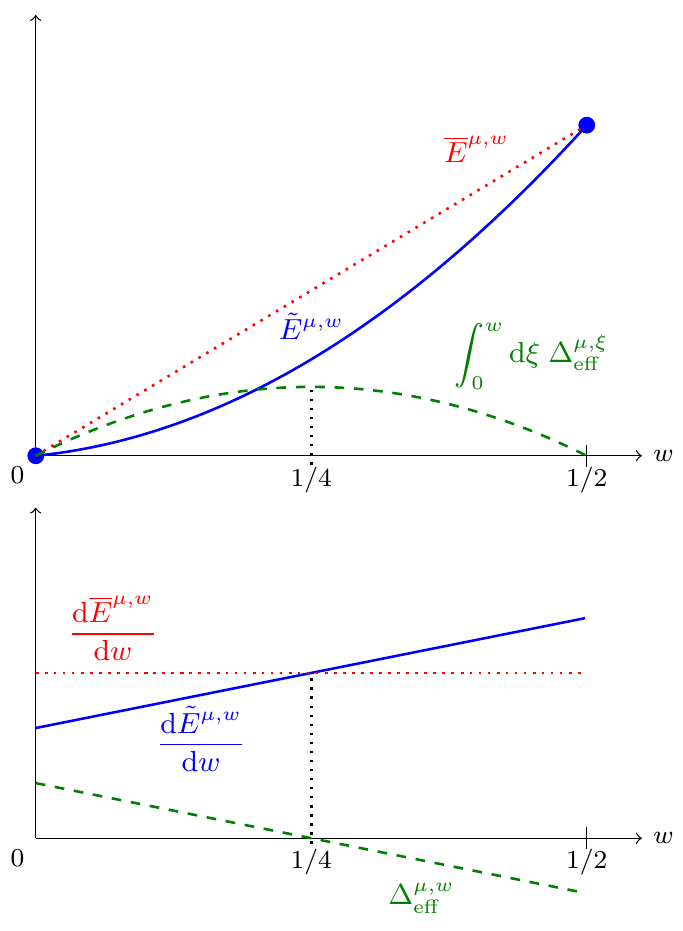}}
\end{tabular}
\end{center}
\end{figure}

\begin{figure}
\caption{\label{Fig:He_w_curves} Senjean et al, Phys. Rev. A}

\begin{center}
\begin{tabular}{c}
\resizebox{18cm}{!}{\includegraphics{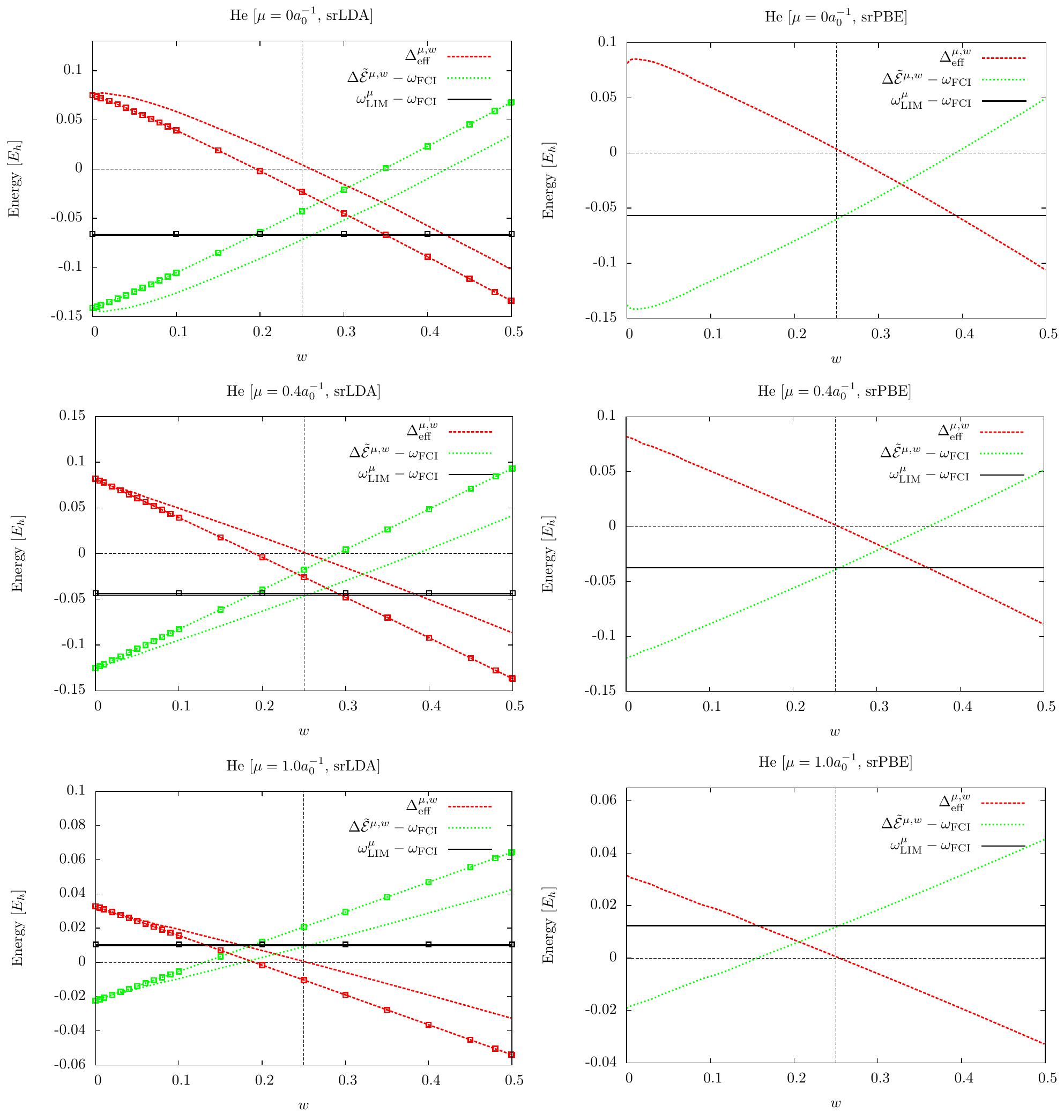}}
\end{tabular}
\end{center}
  \end{figure}

\begin{figure}
\caption{\label{Fig:Be_and_HeHplus_w_curves} Senjean et al, Phys. Rev. A}
\begin{center}
\begin{tabular}{c}
\hspace{-1cm}
\resizebox{18cm}{!}{\includegraphics{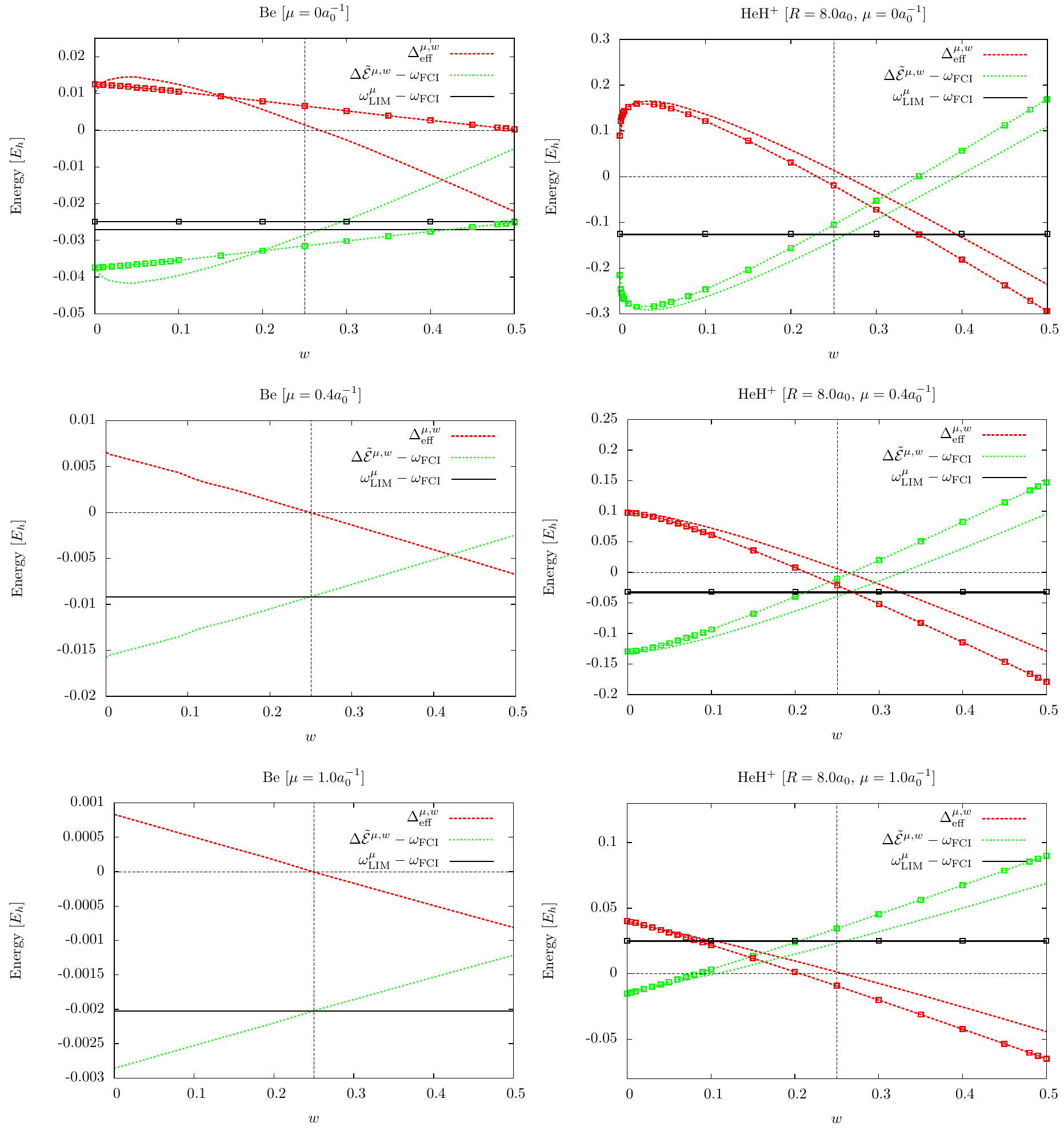}}
\end{tabular}
\end{center}
  \end{figure}

\begin{figure}
\caption{\label{Fig:compar_auxXE_HeBe_HeHplus} Senjean et al, Phys. Rev. A}
\begin{center}
\begin{tabular}{c}
\resizebox{15cm}{!}{\includegraphics{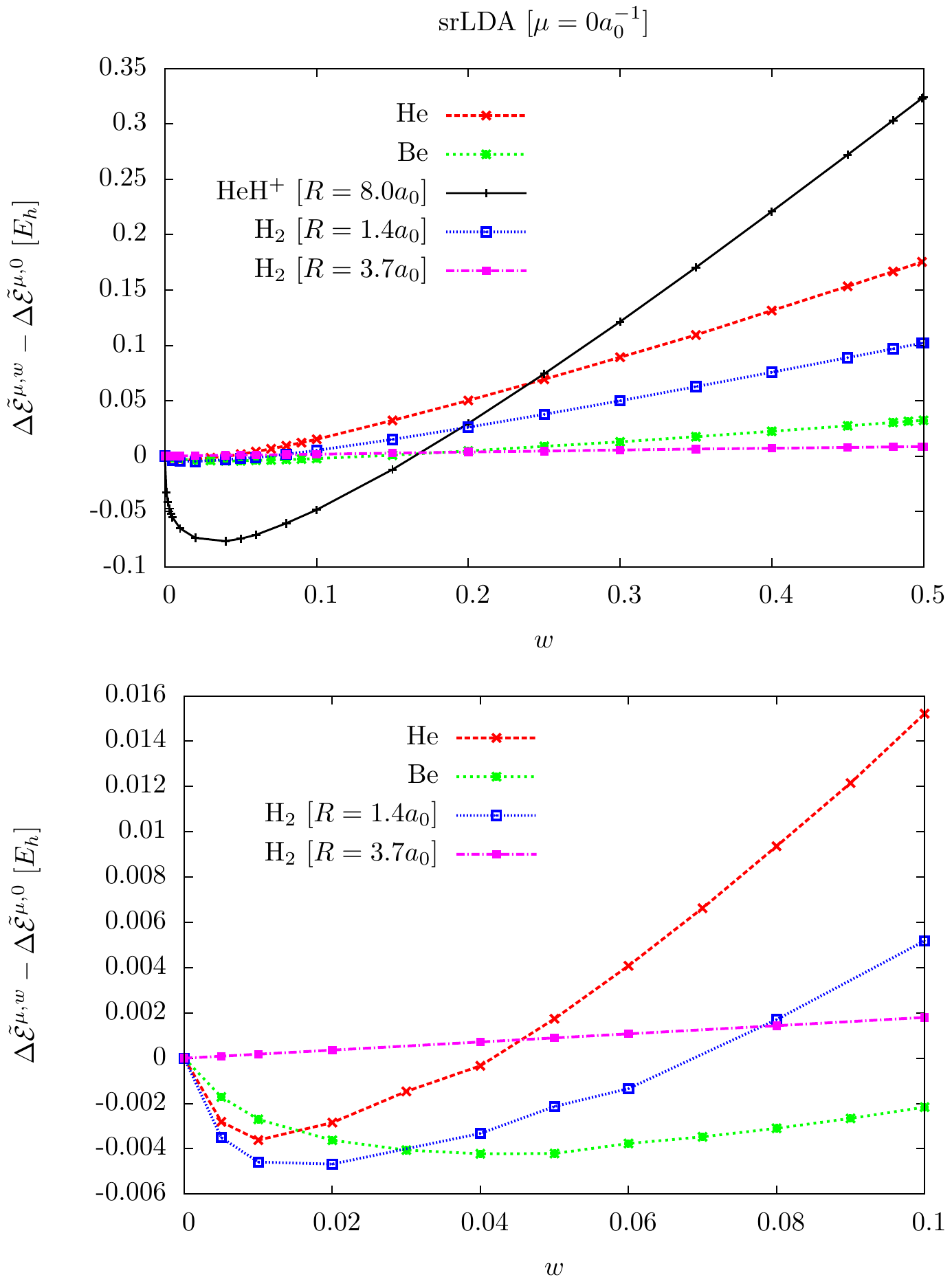}}
\end{tabular}
\end{center}
  \end{figure}

\begin{figure}
\caption{\label{Fig:H2_w_curves} Senjean et al, Phys. Rev. A}
\begin{center}
\begin{tabular}{c}
\hspace{-1cm}
\resizebox{18cm}{!}{\includegraphics{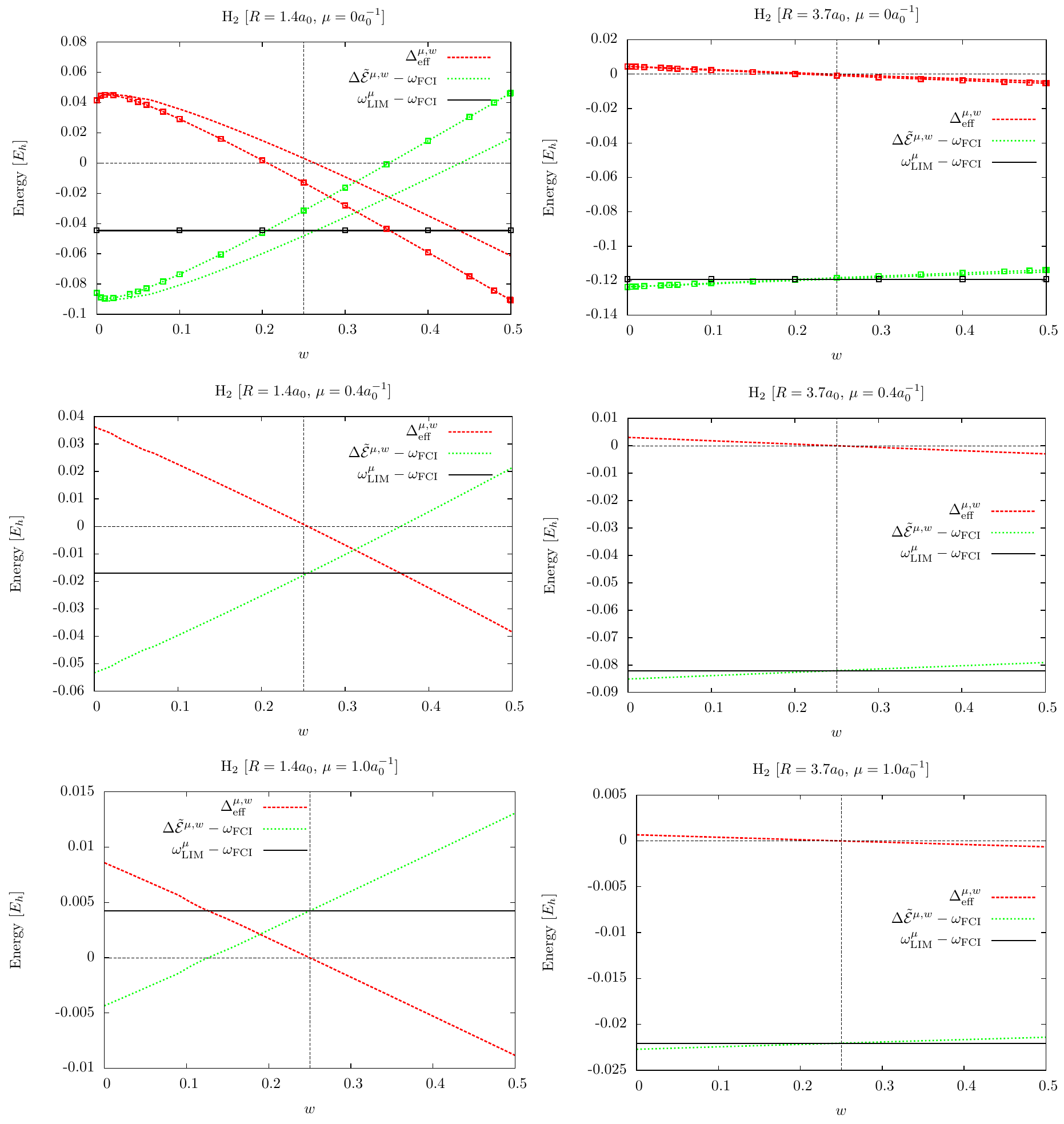}}
\end{tabular}
\end{center}
  \end{figure}

\begin{figure}
\caption{\label{Fig:singles_mu_curves} Senjean et al, Phys. Rev. A}
\begin{center}
\begin{tabular}{c}
\hspace{-1cm}
\resizebox{18cm}{!}{\includegraphics{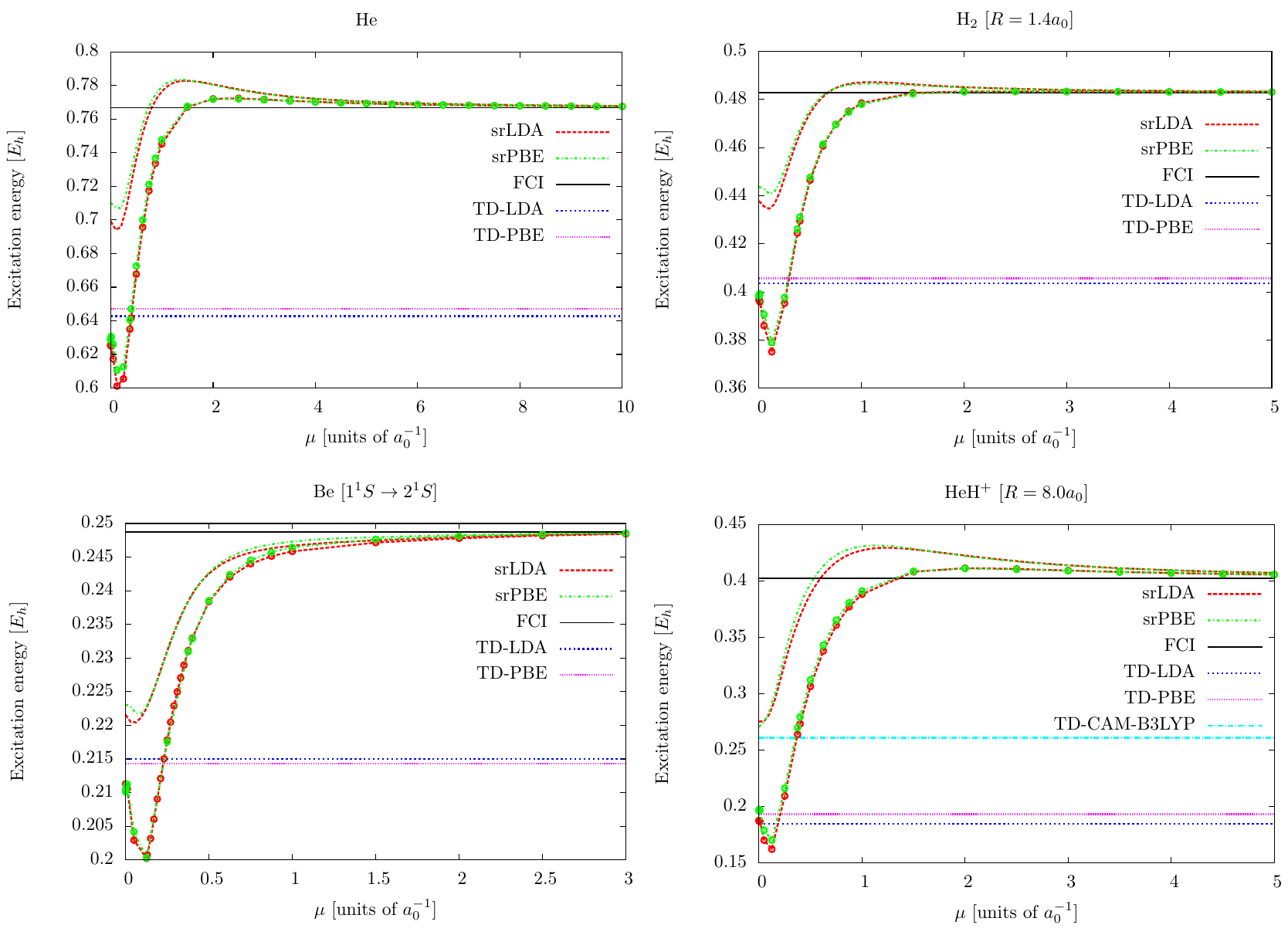}}
\end{tabular}
\end{center}
\end{figure}

\begin{figure}
\caption{\label{Fig:LIM_2ble_XE} Senjean et al, Phys. Rev. A}
\begin{center}
\begin{tabular}{c}
\hspace{-1cm}
\resizebox{15cm}{!}{\includegraphics{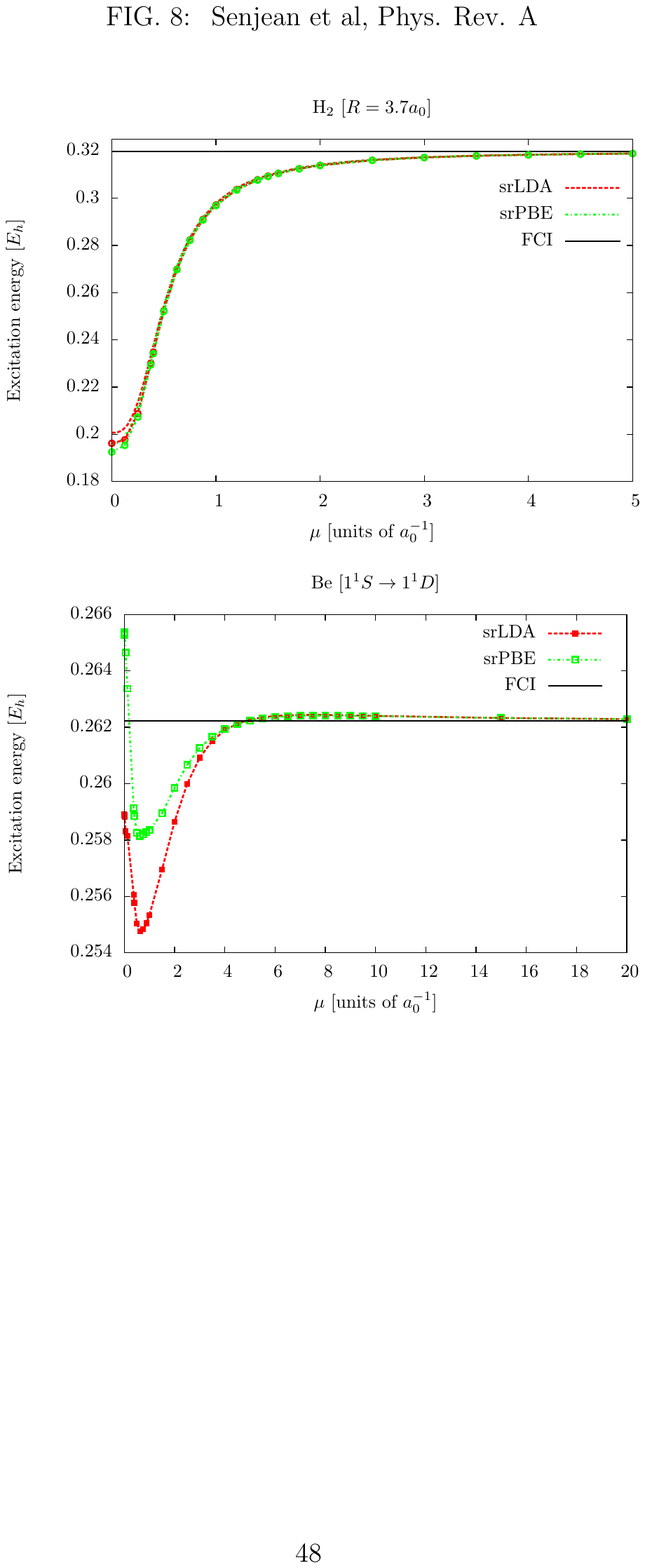}}
\end{tabular}
\end{center}
  \end{figure}

\begin{description}

\item[Figure \ref{Fig:curvature_in_He}] (Color online) 
Range-separated ensemble energy obtained for He at the WIDFA level when varying the
ensemble weight $w$ for $\mu=0$ and $1.0a_0^{-1}$. Comparison is made
with the linear interpolation method (LIM) for $\mu=0a_0^{-1}$ and FCI. The
ensemble contains both $1^1S$ and $2^1S$ states. The srLDA functional
has been used.  

\item[Figure \ref{Fig:curvature_corr_illustration}] (Color online) Schematic 
representation of the linear interpolation method. Ensemble energies and their
first-order derivatives are shown in the top and bottom panels,
respectively. See text for further details. 

\item[Figure \ref{Fig:He_w_curves}] (Color online) Effective DD
($\Delta_{\rm eff}^{\mu,w}$), auxiliary
($\Delta\tilde{\mathcal{E}}^{\mu,w}$) and LIM ($\omega^{\mu}_{\rm LIM}$)
excitation energies associated with the excitation $1^1S\rightarrow
2^1S$ in He. Results are shown for $\mu=0$, 0.4 and 1.0 $a_0^{-1}$ with
the srLDA (left-hand panels) and srPBE (right-hand panels) functionals when varying the ensemble weight $w$.
Comparison is made with the FCI excitation energy $\omega_{\rm
{FCI}}=0.7668$ $E_h$. Empty squares are used for showing non-self-consistent results. 

\item[Figure \ref{Fig:Be_and_HeHplus_w_curves}] (Color online) Effective DD
($\Delta_{\rm eff}^{\mu,w}$), auxiliary
($\Delta\tilde{\mathcal{E}}^{\mu,w}$) and LIM ($\omega^{\mu}_{\rm LIM}$)
excitation energies associated with the excitations $1^1S\rightarrow
2^1S$ in Be (left-hand panels) and $1^1\Sigma^+\rightarrow 2^1\Sigma^+$
in the stretched HeH$^+$ molecule (right-hand panels). Results are shown
for $\mu=0$, 0.4 and 1.0$a_0^{-1}$ with the srLDA functional when
varying the ensemble weight $w$. Comparison is made with the FCI
excitation energies ($\omega_{\rm {FCI}}=0.2487E_h$ for Be and 
$\omega_{\rm {FCI}}=0.4024E_h$ for HeH$^+$). Empty squares are used for showing non-self-consistent results.

\item[Figure \ref{Fig:compar_auxXE_HeBe_HeHplus}] (Color online) 
Auxiliary excitation energies obtained with $\mu=0a_0^{-1}$ and the
srLDA functional (that is equivalent to GOK-LDA) when varying the
ensemble weight $w$ in the various systems considered in this work.
See text for further details. Excitation energies are shifted by their
values at $w=0$ for ease of comparison. A zoom is made on the $0\leq
w\leq 0.1$ region in the bottom panel. 
 
\item[Figure \ref{Fig:H2_w_curves}] (Color online) Effective DD
($\Delta_{\rm eff}^{\mu,w}$), auxiliary
($\Delta\tilde{\mathcal{E}}^{\mu,w}$) and LIM ($\omega^{\mu}_{\rm LIM}$)
excitation energies associated with the excitation $1^1\Sigma^+_g\rightarrow
2^1\Sigma^+_g$ in H$_2$ at equilibrium (left-hand panels) and 
in the stretched geometry (right-hand panels). Results are shown
for $\mu=0$, 0.4 and 1.0$a_0^{-1}$ with the srLDA functional when
varying the ensemble weight $w$. Comparison is made with the FCI
excitation energies ($\omega_{\rm {FCI}}=0.4828E_h$ at equilibrium and 
$\omega_{\rm {FCI}}=0.3198E_h$ in the stretched geometry). Empty squares
are used for showing non-self-consistent results.  

\item[Figure \ref{Fig:singles_mu_curves}] (Color online) 
LIM excitation energies obtained for the single excitations discussed in
this work with srLDA and srPBE functionals
when varying the range-separation parameter $\mu$. Comparison is made
with standard TD-DFT and FCI. For analysis purposes, auxiliary
excitation energies obtained from the ground-state density ($w=0$) are
shown (curves with empty circles). 

\item[Figure \ref{Fig:LIM_2ble_XE}] (Color online) 
LIM excitation energies calculated for the doubly-excited 
$2^1\Sigma^+_g$ state in the stretched H$_2$ molecule (top panel) 
and $1^1D$ state
in Be (bottom panel)
when
varying the range-separation parameter $\mu$ with srLDA and srPBE
functionals. Comparison is made with FCI. For H$_2$, 
auxiliary excitation energies obtained from the ground-state density ($w=0$) are
shown (curves with empty circles) for comparison. 
\end{description}

\clearpage


\clearpage



\clearpage



\clearpage


\clearpage


\clearpage


\clearpage



\clearpage


\clearpage

\end{document}